

\hsize = 7.0in
\hoffset = -.25in
\vsize = 9.5in
\voffset = -0.3in
\tolerance 500


\font\twelverm = cmr12
\font\twelvei  = cmmi12
\font\twelvesy = cmsy10 scaled\magstep1
\font\twelvebf = cmbx12
\font\twelveit = cmti12
\font\twelvesl = cmsl12
\font\twelvett = cmtt12

\font\tenrm = cmr10
\font\tenex = cmex10

\font\ninerm =  cmr9
\font\ninei  = cmmi9
\font\ninesy = cmsy9
\font\ninebf = cmbx9

\font\sixrm=cmr6
\font\sixi =cmmi6
\font\sixsy=cmsy6
\font\sixbf=cmbx6

\def\tenpoint{\def\rm{\fam0\tenrm}
  \textfont0=\tenrm \scriptfont0=\sevenrm \scriptscriptfont0=\fiverm
  \textfont1=\teni \scriptfont1=\seveni \scriptscriptfont1=\fivei
  \textfont2=\tensy \scriptfont2=\sevensy \scriptscriptfont2=\fivesy
  \textfont3=\tenex \scriptfont3=\tenex \scriptscriptfont3=\tenex
  \def\it{\fam\itfam\tenit} \textfont\itfam=\tenit
  \def\sl{\fam\slfam\tensl} \textfont\slfam=\tensl
  \def\bf{\fam\bffam\tenbf} \textfont\bffam=\tenbf \scriptfont\bffam=\sevenbf
  \scriptscriptfont\bffam=\fivebf  \def\tt{\fam\ttfam\tentt}%
  \textfont\ttfam=\tentt
  \normalbaselineskip=12pt
  \normalbaselines\rm}

\def\twelvept{\def\rm{\fam0\twelverm}
  \textfont0=\twelverm   \scriptfont0=\ninerm    \scriptscriptfont0=\sixrm
  \textfont1=\twelvei    \scriptfont1=\ninei     \scriptscriptfont1=\sixi
  \textfont2=\twelvesy   \scriptfont2=\ninesy    \scriptscriptfont2=\sixsy
  \textfont3=\tenex      \scriptfont3=\tenex     \scriptscriptfont3=\tenex
  \textfont\bffam=\twelvebf   \scriptfont\bffam=\ninebf
  \scriptscriptfont\bffam=\sixbf  \def\bf{\fam\bffam\twelvebf}
  \textfont\itfam=\twelveit       \def\it{\fam\itfam\twelveit}
  \textfont\slfam=\twelvesl       \def\sl{\fam\slfam\twelvesl}
  \textfont\ttfam=\twelvett       \def\tt{\fam\ttfam\twelvett}
  \normalbaselineskip=16pt plus 1pt
  \setbox\strutbox=\hbox{\vrule height11pt depth5pt width0pt}
  \let\sc=\ninerm
  \normalbaselines\rm}

\font\bfmone = cmbx12
\font\ffex=cmsy10

\def\eb#1{\setbox0=\hbox{#1}%
\copy0\kern-\wd0
\kern-.02em\copy0\kern-\wd0
\kern+.02em\raise.02ex\copy0\kern-\wd0
\kern+.02em\box0}

\def\pmb#1{\setbox0=\hbox{$#1$}%
\kern-.02em\copy0\kern-\wd0
\kern.05em\copy0\kern-\wd0
\kern-.025em\raise.0433em\box0}

\def\ll{\left\langle}
\def\rr{\right\rangle}
\def\svec#1{\skew{-2}\vec#1}
\def\lapp{\hbox{${\lower.40ex\hbox{$<$}\atop \raise.20ex\hbox{$\sim$}}$}}
\def\lessim
{~\lower0.6ex\hbox{\vbox{\offinterlineskip\hbox{$<$}\vskip1pt\hbox{$\sim$}}}~}
\def\sub#1{\kern-.1em\lower0.6ex\hbox{${}_{#1}$}}
\def\square{\kern1pt\vbox{\hrule height.6pt\hbox{\vrule width.6pt\hskip3pt
\vbox{\vskip6pt}\hskip3pt\vrule width.6pt}\hrule height.6pt}\kern1pt}
\def\footnoterule{\kern-3pt \hrule width \hsize \kern2.6pt}

\headline={\hfil}
\footline={\ifnum\pageno>0 \hss --\folio-- \hss \else\fi}

\twelvept
\pageno=0
\baselineskip 15.5pt plus 1pt

\centerline{\bf IMPROVED COUPLED CHANNELS AND \eb{$R$}-MATRIX MODELS:}
\smallskip
\centerline{{\bf \eb{$pp$}\ PREDICTIONS TO 1~GeV}
\footnote{*}{This work is supported in part by funds provided by the
U.~S.~Department of Energy (D.O.E.) under contract \#DE-AC02-76ER03069.}}
\vskip 20pt
\centerline{P.~LaFrance}
\vskip 10pt
\centerline{\it D\'epartement de physique}
\centerline{\it Universit\'e du Qu\'ebec \`a Montr\'eal}
\centerline{\it C.~P.~8888, succursale A}
\centerline{\it Montr\'eal, H3C 3P8}
\centerline{\it Qu\'ebec, Canada}
\vskip 20pt
\centerline{E.~L.~Lomon and M.~Aw${}^{\hbox{\ninerm (a)}}$}
\vskip 10pt
\centerline{\it Center for Theoretical Physics}
\centerline{\it Laboratory for Nuclear Science}
\centerline{\it and Department of Physics}
\centerline{\it Massachusetts Institute of Technology}
\centerline{\it Cambridge, Massachusetts\ \ 02139\ \ \ U.S.A.}
\vfill
\centerline{\bf ABSTRACT}
\medskip
The $N\!N$ scattering near inelastic threshold is sensitive to the long-range
diagonal interaction in the produced isobar channel.  Earlier models included
meson exchange potentials in the $N\!N$ sector and connecting that sector to
isobar channels, as well as an $R$-matrix description of short-range quark
effects.  Including the diagonal pion exchange contributions in the
$N\!\Delta$ channels coupled to the very inelastic ${}^1D_2$ and ${}^3F_3$
$N\!N$ channels substantially improves those phase parameters and the
observables $\Delta\sigma\sub{L}$ and $\Delta\sigma\sub{T}$.  The same
improvement is made to the other $I=1$ $N\!N$ channels.  In the ${}^1S_0$
$N\!N$ channel coupling to the $\Delta\Delta$ channel is unusually important
due to the $L=2$ angular momentum barrier in the $N\!\Delta$ channel.  The
pion exchange transition potentials between isobar channels are included in
this partial wave to obtain the correct equilibrium between one- and two-pion
decay channels.  Other improvements to earlier models have been made, in the
specification of isobar channels and the inclusion of decay width effects in
more channels.  A comparison is made with all the $pp$ data for $T_L \le
800$~MeV, producing a very good fit over the whole energy range.
\vfill
\centerline{Submitted to: {\it Physical Review C}}
\vfill
\noindent CTP\#2133\hfill May 1993
\eject

\baselineskip 24pt plus 1pt
\noindent{\bfmone I.\quad Introduction}\medskip\nobreak
In describing the nucleon-nucleon ($N\!N$) interaction beyond the first
150~MeV of kinetic energy in the center of momentum system (nucleon beam
energy $T_L>$ 300~MeV), coupling to isobar channels introduces an important
energy dependence to the real phase even when the imaginary phase remains
small.  Single-pion production begins at $T_L = 280$~MeV, but is relatively
weak until the $\Delta$ production threshold, $T_L = 630$~MeV is approached.
But the effect on the elastic scattering begins below the threshold of
substantial inelasticity because of the effect of virtual intermediate states
of higher mass.  Consequently, a coupled channel system must be used to
understand the $N\!N$ reaction in the whole range $0\le T_L\le 800$~MeV.
Experience has shown$^{1,\,2,\,3}$ that for some $N\!N$ partial waves the
$\Delta\Delta$ and $N\!N^*$(1440) channels are important, in addition to the
low threshold $N\!\Delta$ channel.

The $R$-matrix (or Boundary Condition) model$^{4,\,5}$ incorporates long-range
meson exchange effects in the potential matrix outside the boundary radius
$r_0$.$^1$ The internal dynamics, including the effects of quark degrees of
freedom, is described by a simple boundary condition at $r_0$.  The boundary
condition is a meromorphic function of energy$^4$ whose poles are completely
described by the properties of simple quark configurations.$^{2,\,3,\,6,\,7}$
Only the constant term is not given by the internal model.

Using a potential matrix and a boundary condition matrix the coupling to
isobaric channels is easily incorporated.$^1$ With some computational burden,
but little formal complication, the width of the isobars is included$^8$ by
discretizing the Breit--Wigner mass distribution of the isobar, treating each
mass as a separate channel.  In previous work,$^{3,\,8}$ this has provided a
good quantitative description of the phase parameters of most of the
proton-proton ($pp$) partial waves.  The most important exceptions to this
quantitative agreement have been the detailed shapes of the inelasticity,
$\eta$, in the ${}^1D_2$ and ${}^3F_3$ channels.$^8$ The predicted positions,
strengths and widths of the structure in these channels are in agreement with
experiment, indicating that the structures are due to the effect of the
$N\!\Delta$ threshold and require no significant contribution from quark
substructure (the latter is important at somewhat higher
energies$^{2,\,3,\,6,\,7}$).  In fact this potential and $R$-matrix model has
been the only one to adequately predict the strong onset of inelasticity in
the ${}^3F_3$ channel.  However, because of the dominance of these channels in
the $\Delta\sigma\sub{L}(pp)$ and $\Delta\sigma\sub{T}(pp)$ structures, for
600~MeV $<T_L<800$~MeV,$^9$ the detailed shaped of the inelasticity must be
fairly precise to reproduce the data in the region.

In the model, the $\pi,\eta,\rho$ and $\omega$ meson exchanges (and also
two-pion exchange) are all included in deriving the potential within the
$N\!N$ sector (Fig.~1(a)--(c)).

Because of the indirect effect of the $N\!N$ to isobar channel transition
(Fig.~1(d)--(e)) on the $N\!N$ elastic interaction only the $\pi$ meson
exchange transition potentials are included.  A two-pion exchange transition
potential is only important at distances less than one-third of a pion Compton
wavelength in the $N\!N \to N\!N$ reactions, at which distance it is partly
masked by the effect of the boundary condition.  Nevertheless, a
phenomenological two-pion range transition potential is sometimes included,
which may also substitute for $\rho$ meson exchange.

The diagonal $N\!\Delta$ interaction is one stage further removed than the
transition interaction from influencing the $N\!N$ scattering, and was
entirely represented by the boundary condition in previous work.  However,
when the energy dependence of the effect is strong, as it is near the
$N\!\Delta$ threshold, then the short-range boundary condition is not an
adequate substitute for the long-range diagonal potential. The long-range
interaction shown in Fig.~1(f)--(g) is more important at small momentum
transfers.  In this paper we include these diagonal potentials, improving many
partial wave phases.  In particular, the $\eta$ and $\delta$ in the ${}^1D_2$
and ${}^3F_3$ channels are greatly improved for $T_L>500$~MeV.

In previous work$^{2,\,3}$ there was some difficulty in obtaining a very good
fit to the higher energy ${}^1S_0$ $N\!N$ phases simultaneously with correctly
predicting the experimentally very small two-pion production at
$T_L=800$~MeV.$^{10}$~ The $N\!\Delta$ diagonal potential affects this
situation positively, but the transition between $N\!\Delta$ channels, which
produce one pion, and $\Delta\Delta$ or $N\!N^*$(1440) channels, which produce
two pions, critically affects the pion production multiplicity.  The
long-range one-pion exchange transition potentials between isobar channels are
important to this spectrum at low energies.  We include them here in the
${}^1S_0$ channel and obtain good results for both elastic $N\!N$ scattering
and pion production.

In previous work some isobar channels differing only in spin were combined.
They are now treated as separate channels to more accurately treat the pion
exchange potentials and for future predictions of 3-body final state
distributions.  The effect of isobar width is included in more channels.

In some instances the one-pion exchange (OPE) coefficients for particular
channels were incorrect in previous results.  These corrections will be noted
in the description of each partial wave.  $r_0$ and $g\sub{N\!N\pi}$ have been
chosen so that one value is used throughout.

In Section~II we review critical aspects of the model.  Section~III describes
the model elements and parameters for each $N\!N$ partial wave and displays
the fit to the phase parameters.  In Section~IV all the model partial waves
with $J\le 4$, and the one-pion-exchange amplitude for higher $J$ partial
waves are used to predict the experimental observables.  The excellent results
are displayed with full angular distribution of all spin observables for which
there is data at key energies spanning $T_L\le 800$~MeV.  In addition, we
present excitation curves at $90^\circ$ for these observables and total cross
sections (with and without polarization) over the whole range.  In particular,
we note that the well-known structures$^9$ in $\Delta\sigma\sub{L}(pp)$ and
$\Delta\sigma\sub{T}(pp)$ are well-fitted with the exception of the somewhat
shifted position of the $\Delta\sigma\sub{T}$ peak.  This defect is correlated
with a slightly low minimum value of $\eta\left({}^1D_2\right)$.  The physics
that may account for this will be discussed.

\goodbreak\bigskip
\noindent{\bfmone II.\quad The Hadron Interaction Model}\medskip
\noindent{\bf A.\quad The Schr\"odinger Equation}\medskip\nobreak
For the internucleon distance $r>r_0$, the coupled system wave function is
determined by a homogeneous boundary condition at $r_0$ (to be discussed
later) and a coupled channel Schr\"odinger potential with a meson exchange
potential matrix.  Using the strong interaction symmetries we assume
conservation of total angular momentum, $J$, isospin $I$ and parity.  As is
well-known this implies total spin, $S$, conservation in the $N\!N$ system
(neglecting the $n-p$ mass difference).  Consequently, the coupled system
contains only one $(L=J)$ or two ($L=J\pm 1$, $S=1$) $N\!N$ partial waves, and
may have as many $N\!\Delta$, $\Delta\Delta$ or $N\!N^*$(1440) partial waves
as have the same $J,I$ and parity.  Because of the width of the isobars each
isobar channel partial wave is treated as many channels,$^8$ with the mass
distribution discretized and weighted by the Breit--Wigner distribution.  For
each channel, $i$, the eigenvalue is determined by the relativistic value of
the relative momentum $k_i$ in that channel, where the total energy $W
=\sqrt{{M^a_i}^2 + k^2_i} + \sqrt{{M^b_i}^2 + k^2_i}$ with $M^a_i$ and $M^b_i$
being the pair of baryon masses in channel $i$.  The system of Schr\"odinger
equations is described in detail in Ref.~[8].

\goodbreak\bigskip
\noindent{\bf B.\quad The Potential in the \eb{$N\!N$}\ Sector}\medskip\nobreak
Within the $N\!N$ sector, the potential matrix (diagonal and tensor coupling
elements) is determined by the one-boson exchange of $\pi$, $\eta$, $\rho$ and
$\omega$ mesons (Fig.~1(a)) and two-pion exchange (Fig.~1(b)--(c)) in the
adiabatic limit.  These potentials, with the exception of the spin-orbit term,
are described in Ref.~[5].  In particular, the form of the two-pion potential
is detailed in that reference including the ``pair suppression'' factor and
the Breuckner--Watson vs. Taketani--Machida--Ohnuma ambiguity which arises in
the adiabatic limit.  All the coupling constants (with the exception noted
below), the meson masses and the two parameters which arise in two-pion
exchange potential are as in Ref.~[5]. The exception is a small change in the
value of the pseudoscalar coupling $g_{N\!N\pi}$.  As a result of the larger
core separation radius now used, and the coupling to isobars, the value now
used is $g^2_{N\!N\pi}/4\pi=14.40$ (in place of the value 14.94 in Ref.~[5]).

The spin-orbit potential, ${\svec L} \cdot {\svec S}\, V_{LS}$, was ignored in
Ref.~[5] because of its short range.  The tail of this potential beyond $r_0$
has been found to be of some importance to the $L\not=0$ partial waves.  The
sources of this term are two-pion, $\rho$ and $\omega$ exchange, so that it
has contributions of different range, form (the two-pion part is non-Yukawa)
and sign.  Because of the relatively small overall effect of this part of the
potential, the radial form has been simplified by fitting a single Yukawa
potential to the complete, relativistic result of Ref.~[11].  In the $I=1$
states this results in $$V_{LS} = - 7 {e^{-5m_\pi r}\over r}\ \ .\eqno(2.1)$$
It is smaller in the $I=0$ states and we neglect it.

\goodbreak\bigskip
\noindent{\bf C.\quad The Transition Potentials}\medskip\nobreak

The off-diagonal components of the potential matrix connecting to isobar
channels are obtained from one-pion exchange as shown in Fig.~1(d)--(e).  The
general form is$^{12}$
$$
V_T(r) = {m_\pi\over 3}\ {f_1 f_2\over 4\pi} {\svec T}_1 \cdot {\svec T}_2
\left[ {\svec S}_1 \cdot {\svec S}_2 v_0(r) + S^T_{12} v_2 (r)
\right]\eqno(2.2)
$$
where the $f_i$ are the coupling constants at vertex $i$, the
${\svec S}_i\left( {\svec T}_i\right)$ are spin (isospin) or
transition spin (isospin) operators at vertex $i$ that will be defined below,
the tensor operator is
$$
S^T_{12} = 3\left( {\svec S}_1\cdot \hat r\right)
\left( {\svec S}_2\cdot \hat r\right) - {\svec S}_1\cdot {\svec S}_2\ \ ,
\eqno(2.3)
$$
$$
v_0(r) = {e^{-m_\pi r}\over m_\pi r}\qquad\hbox{and}\qquad v_2(r) = \left( 1 +
{3\over m_\pi r} + {3\over \left( m_\pi r\right)^2}\right) v_0(r) \ \
.\eqno(2.4)
$$

The matrix elements of the above operators are expressible as$^{12,\,13}$
$$\eqalign{
&\ll S'_1S'_2S'L'J\left|{\svec S}_1 \cdot {\svec S}_2\right| S_1\, S_2\,
SLJ\rr \cr &= \left( -1\right)^{S_1+S'_2 + S}\delta_{SS'}\delta_{LL'}
\left\{ \matrix{
S & S'_2 & S'_1 \cr\noalign{\vskip 0.2cm}
1 & S_1 & S_2 \cr}\right\} \left( S'_1 \left\| {\svec S}_1 \right\|
S_1\right) \left( S'_2 \left\| {\svec S}_2\right\| S_2\right)\ \
,\cr}\eqno(2.5)$$
the analogous expression for ${\svec T}_1\cdot {\svec T}_2$ and
$$\eqalign{&\ll S'_1S'_2 S' L'J\left| S^T_{12}\right| S_1 S_2 SLJ\rr
= \left( -1\right)^{S'+J} \left[ 30(2L+1) (2L'+1) (2S+1)
(2S'+1)\right]^{1/2}\cr
&\times
\left\{\matrix{ J & S' & L'\cr\noalign{\vskip 0.2cm} 2 & L & S \cr}\right\}
\left( \matrix{ L' & 2 & L\cr\noalign{\vskip 0.2cm} 0 & 0 & 0 \cr}\right)
\left\{ \matrix{ S'_1 & S_1 & 1 \cr\noalign{\vskip 0.2cm} S'_2 & S_2 &
1 \cr S' & S & 2 \cr}\right\}
\left( S'_1\left\| {\svec S}_1\right\| S_1\right) \left( S'_2
\left\| {\svec S}_2 \right\| S_2\right) \cr}\eqno(2.6)$$
where the standard notation is used for the $3j$ (Wigner), $6j$ (Racah) and
$9j$ coefficients.  The Wigner--Eckart theorem requires$^{13}$
$$\ll S'm' \left| S^M\right| S,m\rr = N\ll S'm'|1M,Sm\rr\eqno(2.7)$$
where the normalization $N$ is determined by convention and the $\ll S'm'
| JM, Sm\rr$ are Clebsch--Gordan coefficients.  In the Edmonds
convention$^{14}$
$N=1$ for the transition spin operators.  For those cases diagonal in spin
magnitude, because the Pauli spin operators and the equivalent spin-3/2
operator are twice the standard angular momentum operator normalization
\hbox{${\svec J}^2 = J(J+1)$},
$N=-2\sqrt{S(S+1)}$.  For $N\!N\pi$, $N\!N^*\pi$, $N\!\Delta \pi$ and
$\Delta\Delta\pi$ vertices this requires
$$\eqalignno{
\left( {1\over 2} \left\| {\svec\sigma}\right\| {1\over 2}\right)
&= \sqrt{6}\ \ ,
&(2.8\hbox{a}) \cr\noalign{\vskip 0.2cm}
\left( {3\over 2} \left\|{\svec S}_{1/2,3/2} \right\| {1\over 2}\right) &=
-\left( {1\over 2} \left\| {\svec S}_{3/2,1/2}\right\| {3\over 2}\right) =2
&(2.8\hbox{b}) \cr\noalign{\hbox{and}}
\left( {3\over 2}\left\| {\svec\Sigma}\right\| {3\over 2}\right) &= 2\sqrt{15}
&(2.8\hbox{c}) \cr}$$
respectively, with analogous results for the isospin operators.

With these normalizations of the operators we have
$$f\sub{N\!N\pi} = {m_\pi\over 2M} g\sub{N\!N\pi}\eqno(2.9)$$
where in $pp$ scattering, $m_\pi$ is the neutral pion mass and
$M$ is the proton mass. As discussed in Ref.~[1], we obtain
$f_{N\!\Delta\pi}$ from the decay width of the $\Delta$, giving
$f^2_{N\!\Delta\pi}/4\pi = 0.35$
which is 50\% larger than the quark model
value but only 5\% smaller than the result of a strong coupling model.$^{12}$
Similarly, we obtain${}^1$ $f_{N\!N^*\pi}$
from the $N^*(1440)$ decay to a single pion,
giving $f^2_{N\!N^*\pi} / 4\pi = 0.015$.
Lacking other information we have used the consistent quark and strong
coupling model result$^{12}$
$$\eqalign{f_{\Delta\Delta\pi} &= {1\over 5} f_{N\!N\pi}\ \ ,
\cr\noalign{\vskip 0.2cm}
f^2_{\Delta\Delta\pi} / 4\pi &= 0.003 \ \ .\cr}\eqno(2.10)$$

\goodbreak\bigskip
\noindent{\bf D.\quad The Diagonal \eb{$N\!\Delta$}\ Potential}\medskip\nobreak
The equations and matrix elements for the diagonal $N\!\Delta$ potential (see
Fig.~1(f)--(g)) are of the same form as above in Section~IIC.  Here we discuss
the implications of the fact that there are contributions from two distinct
Feynman diagrams: the direct interactions, Fig.~1(f), and the exchange
interaction, Fig.~1(g).  To obtain its contribution to
$V_{N\!\Delta,N\!\Delta}$, the exchange amplitude must be multiplied by
$(-1)^{L+S+T+1}$.

First we note that the coupling constant product for the direct interaction is
$f_{N\!N\pi} f_{\Delta\Delta\pi} / 4\pi = 0.0155$, while that for the exchange
interaction is $f^2_{N\!\Delta\pi} / 4\pi =0.351$.  However, that apparent
dominance of the exchange interaction is largely reversed by the difference of
the spin matrix elements.  In the isotriplet state the matrix element of
${\svec\tau}\cdot {\svec T}$ is $-5$, but the matrix element of ${\svec
T}_{1/2,3/2}\cdot {\svec T}_{3/2,1/2}$ is $-{1\over 3}$.  In addition, the
spin-spin and tensor matrix elements are in most states larger for the direct
interaction.  The result is that the two contributions are of the same order,
and as they sometimes differ in sign, the strength and sign of the long-range
diagonal $N\!\Delta$ interaction is very sensitive to the channel involved.
One may then hope to determine the effect of the diagonal coupling by
comparing the effect in different channels.  As we shall see, this is in fact
the case in the ${}^1D_2$ and ${}^3F_3$ channels, where strong inelasticity
enhances the sensitivity to the diagonal components.

\goodbreak\bigskip
\noindent{\bf E.\quad The \eb{$R$}-Matrix Boundary Condition}\nobreak
\medskip\nobreak
As shown in Ref.~[4], the boundary condition at $r_0$ is of the general form
$(\alpha,\beta$ are channel indices)
$$R_0 {d\psi^W_\alpha\over dr_0}
= \sum_\beta f_{\alpha\beta} (W) \psi^W_\beta(r)
\eqno(2.11)$$
where $W$ is the total barycentric energy and the $f_{\alpha\beta}(W)$ are
meromorphic functions of $W$ with poles of positive residue on the real axis.
As discussed in Refs.~[4] and [15], these poles are determined by the
complete set of states in the interior satisfying $\psi_{\rm int} (r_0) = 0$.
 The pole position is given by the eigenvalue $W_i$
of the corresponding interior
state, and the residue by the derivatives of the interior wavefunctions at
$r_0$.  It was shown in Ref.~[16] that the residues can be re-expressed in
terms of the rate of change of eigenvalue with $r_0$
$$\rho^i_{\alpha\beta} = - r_0 {\partial W_i\over \partial r_0} \xi^i_\alpha
\xi^i_\beta\eqno(2.12)$$
where the $\xi^i_\alpha$ are the fractional parentage coefficients of the
channel $\alpha$ in the interior state $i$.

We then have
$$f_{\alpha\beta}(W) = f^{\,0}_{\alpha\beta} + \sum_i
{\rho^i_{\alpha\beta}\over W
- W_i}\ \ .\eqno(2.13)$$

The fractional parentage coefficients relevant to our models are those of the
$\left( {}^1 S_{1/2}\right)^6$ quark configuration which overlaps with the
$N\!N\left({}^1S_0\right)$, $\Delta\Delta\left( {}^1S_0\right)$,
$N\!\Delta\left( {}^5S_2\right)$ and $\Delta\Delta\left( {}^5 S_2\right)$
channels.  They are $\xi^{0^+}_{N\!N} = {1\over 3}$, $\xi^{0^+}_{\Delta\Delta}
= {2\over \sqrt{45}}$, $\xi^{2^+}_{N\!\Delta} = {1\over \sqrt{6}}$ and
$\xi^{2^+}_{\Delta\Delta} = {1\over \sqrt{30}}$.

References [2] and [3] discuss the theoretical constraints on $r_0$, and the
experimental determination $r_0 = 1.05$~fm when the quark interior is
described by the Cloudy Bag Model (CBM) parameters.  In the CBM, in contrast
to the MIT Bag Model, the experimentally determined $r_0$ is consistent with
the theoretical constraints.

\goodbreak\bigskip
\hangindent=27pt\hangafter=1
\noindent{\bfmone III. \enskip
Isobar Channels, Meson Exchange Potentials,
Quark Configurations and Parameters
of the \eb{$pp$}\ Partial Waves}\medskip\nobreak
For $J>4$ the amplitude used is that of OPE and the Coulomb force.  Below we
give the details of the model for each $N\!N$, $I=1$ partial wave with
$J\le 4$.

There are several properties of the model that are common to all the partial
wave models. The $R$-matrix separation radius, $r_0$, is determined by
requiring that the $\left[q\left( 1S_{1/2}\right)\right]^6$ configuration
energy, as determined by the CBM, be the energy for which the inner zero of
the external wave function is at $r_0$.  The external wave function is
determined by the potential matrix and the $S$-matrix.  The data determines
the latter for $T_L<1$~GeV.  For the ${}^1S_0$ and ${}^3S_1$--${}^3D_1$ states
the CBM and the data require$^2$ $r_0 = 1.05$~fm.  The same value is indicated
for the ${}^1D_2$ state.$^3$ This value is about 20\% smaller than the value
of $r$ corresponding to the equilibrium radius of these six-quark bags, which
satisfies the $R$-matrix method requirement that $r_0$ be within the region of
asymptotic freedom.$^2$ We use $r_0=1.05$~fm for all partial waves.

The coupling constants used at the meson-baryon vertices are those given in
Refs.~[1] and [5], with the exception that we use $g^2_{N\!N\pi}/4\pi=14.40$
in all partial waves.  This change is related to the larger $r_0$ now being
used.  One other consistent change from the earlier published work is an extra
factor $\sqrt{2}$ in the coefficients of the $N\!N \to N\!\Delta$ potentials.
This is required by contribution of the $N\!N\to \Delta N$ diagram, which is
of the same strength as that of the $N\!N\to N\!\Delta$ diagram.

Table~I presents the values of all the isospin-isospin, spin-spin and tensor
matrix elements that are needed for the determination of the OPE potentials
between $N\!N$ and isobar channels or between isobar channels, for isobar
channels with one or two $\Delta$'s.  The specific numerical coefficients are
given in the subsection for each partial wave.

We note that the $f$-matrix parameters determined by fitting the data depend
slightly on the number and distribution of the discretized mass spectrum of
the isobars. We have used 17 channels for each isobar whose width is not
neglected.  Other aspects of the distribution are discussed in Ref.~[8].

\goodbreak\bigskip
\noindent{\bf A.\quad The \eb{$N\!N\left({}^1S_0\right)$}\ System}
\medskip\nobreak

\midinsert
{\tenpoint
\centerline{\bf Table I.}
\smallskip
\centerline{Spin-spin and tensor matrix element for
$N\!N$-$N\!\Delta$,
$N\!N$-$\Delta\Delta$,
$N\!\Delta$-$N\!\Delta$ and
$N\!\Delta$-$\Delta\Delta$ transitions.}
\smallskip
\centerline{$\alpha$ and $\alpha'$ represent
the sets of quantum numbers in the first two columns.}
\smallskip

\newbox\evstrutbox
\setbox\evstrutbox=\hbox{\vrule height12pt depth6pt width0pt}
\def\evstrut{\relax\ifmmode\copy\evstrutbox\else\unhcopy\evstrutbox\fi}

\def\h{{1\over2}}
\def\t{{3\over2}}
\def\d{&&${\tt "}$&&}

$$\vbox{\offinterlineskip
\hrule\halign{\tabskip=0.0in
\evstrut\vrule#&
\hfil\quad$#$\enskip\hfil&
\hfil\enskip$#$\enskip\hfil&
\hfil\enskip$#$\enskip\hfil&
\hfil\enskip$#$\enskip\hfil&
\hfil\enskip$#$\quad\hfil&
\vrule#&
\hfil\quad$#$\enskip\hfil&
\hfil\enskip$#$\enskip\hfil&
\hfil\enskip$#$\enskip\hfil&
\hfil\enskip$#$\enskip\hfil&
\hfil\enskip$#$\quad\hfil&
\vrule#&
\hfil\quad$#$\quad\hfil&
\vrule#&
\hfil\quad$#$\quad\hfil&
\vrule#\cr
\vrule height16pt depth10pt width0pt
&J&L&S_1&S_2&S&&J&L'&S'_1&S'_2&S'&&
\langle\,\alpha'\,| \, \vec{S_1}\cdot\vec{S_2} \, |\,\alpha\,\rangle&&
\langle\,\alpha'\,| \,      S_{1 2}^{\,T}      \, |\,\alpha\,\rangle&\cr
\noalign{\hrule}
\vrule height15pt depth6pt width0pt
& 0&0&\h&\h&0 && 0&2&\h&\t&2 && 0 &&+\sqrt{6}&\cr
& \d          && 0&0&\t&\t&0 && -\sqrt{2} && 0 &\cr
& \d          && 0&2&\t&\t&2 && 0 && \sqrt{2} &\cr
& 0&2&\h&\t&2 && 0&0&\t&\t&0 && 0 && +\sqrt{3} &\cr
& \d          && 0&2&\t&\t&2 &&-2\sqrt{3}&&\sqrt{3} &\cr
& 1&0&\h&\t&1 && 1&0&\t&\t&1 &&-2\sqrt{5/3}&&$---$&\cr
& 0&2&\h&\t&2 && 0&2&\h&\t&2 && 3 && -6  &\cr
& \d          && 0&2&\t&\h&2 && +1 && 1 &\cr
& 0&1&\h&\h&1 && 0&1&\h&\t&1 && +\sqrt{8/3} && +\sqrt{2/3} &\cr
& \d          && 0&1&\t&\t&1 && -\sqrt{10}/3 && {2\over3} \sqrt{2/5} &\cr
& 0&1&\h&\t&1 && 0&1&\h&\t&1 && -5 && 2  &\cr
& \d          && 0&1&\t&\h&1 && + {1\over3} && {5\over3} &\cr
& 1&1&\h&\h&1 && 1&1&\h&\t&1 && +\sqrt{8/3} && -\sqrt{1/6} &\cr
& \d          && 1&1&\h&\t&2 && 0 && 3\sqrt{3/10} &\cr
& \d          && 1&3&\h&\t&2 && 0 && 3/\sqrt{5} &\cr
& \d          && 1&1&\t&\t&1 && -\sqrt{10}/3 && -{1\over3} \sqrt{2/5} &\cr
& 1&1&\h&\t&1 && 1&1&\h&\t&1 && -5 && -1 &\cr
& \d          && 1&1&\t&\h&1 && + {1\over3} && -{5\over6} &\cr
& 1&1&\h&\t&2 && 1&1&\h&\t&2 && 3 && -{21\over5} &\cr
& \d          && 1&1&\t&\h&2 && +1 && {7\over10} &\cr
& 1&3&\h&\t&2 && 1&3&\h&\t&2 && 3 && -{24\over5}  &\cr
& \d          && 1&3&\t&\h&2 && +1 && {4\over5} &\cr
& 2&1&\h&\h&1 && 2&1&\h&\t&1 && + \sqrt{8/3} && +1/(5\sqrt{6}) &\cr
& \d          && 2&1&\h&\t&1 && 0 && -{3\over5} \sqrt{3/2} &\cr
& \d          && 2&3&\h&\t&1 && 0 && -{3\over5} &\cr
& \d          && 2&3&\h&\t&2 && 0 && {3\over5} \sqrt{6} &\cr
& 2&3&\h&\h&1 && 2&1&\h&\t&1 && 0 && -{3\over5} &\cr
& \d          && 2&1&\h&\t&2 && 0 && -{3\over5} &\cr
& \d          && 2&3&\h&\t&1 && + \sqrt{8/3} && {2\over5} \sqrt{2/3} &\cr
& \d          && 2&3&\h&\t&2 && 0 && {6\over5} &\cr
& 2&1&\h&\t&1 && 2&1&\h&\t&1 && -5 && {1\over5} &\cr
\vrule height12pt depth9pt width0pt  
& \d          && 2&1&\t&\h&1 && + {1\over3} && 1/6 &\cr
}\hrule}$$
\vfill\eject}
\endinsert

\midinsert
{\tenpoint
\centerline{\bf Table I. contd.}
\smallskip
\newbox\evstrutbox
\setbox\evstrutbox=\hbox{\vrule height12pt depth6pt width0pt}
\def\evstrut{\relax\ifmmode\copy\evstrutbox\else\unhcopy\evstrutbox\fi}
\def\h{{1\over2}}
\def\t{{3\over2}}
\def\d{&&${\tt "}$&&}
$$\vbox{\offinterlineskip
\hrule\halign{\tabskip=0.0in
\evstrut\vrule#&
\hfil\quad$#$\enskip\hfil&
\hfil\enskip$#$\enskip\hfil&
\hfil\enskip$#$\enskip\hfil&
\hfil\enskip$#$\enskip\hfil&
\hfil\enskip$#$\quad\hfil&
\vrule#&
\hfil\quad$#$\enskip\hfil&
\hfil\enskip$#$\enskip\hfil&
\hfil\enskip$#$\enskip\hfil&
\hfil\enskip$#$\enskip\hfil&
\hfil\enskip$#$\quad\hfil&
\vrule#&
\hfil\quad$#$\quad\hfil&
\vrule#&
\hfil\quad$#$\quad\hfil&
\vrule#\cr
\vrule height16pt depth10pt width0pt
&J&L&S_1&S_2&S &&J&L'&S'_1&S'_2&S'&&
\langle\,\alpha'\,|\,\vec{S_1}\cdot\vec{S_2}\,|\,\alpha\,\rangle&&
\langle\,\alpha'\,|\,S_{1 2}^{\,T}\,|\,\alpha\,\rangle&\cr
\noalign{\hrule}
\vrule height15pt depth6pt width0pt
& 2&1&\h&\t&2 && 2&1&\h&\t&2 && 3 && {21\over5} &\cr
& \d          && 2&1&\t&\h&2 && +1 && -{7\over10} &\cr
& 2&3&\h&\t&1 && 2&3&\h&\t&1 && -5 && {4\over5} &\cr
& \d          && 2&3&\t&\h&1 && +{1\over3} && {2\over3} &\cr
& 2&3&\h&\t&2 && 2&3&\h&\t&2 && 3 && -{6\over5} &\cr
& \d          && 2&2&\t&\h&2 && +1 && {1\over5} &\cr
& 2&2&\h&\h&0 && 2&0&\h&\t&2 && 0 && \sqrt{6/5} &\cr
& \d          && 2&2&\h&\t&2 && 0 && -2\sqrt{3} &\cr
& \d          && 2&0&\t&\t&2 && 0 && \sqrt{2/5} &\cr
& 2&0&\h&\t&2 && 2&0&\h&\t&2 && 3 && 0 &\cr
& \d          && 2&0&\t&\h&2 && +1 && 0 &\cr
& 2&2&\h&\t&2 && 2&2&\h&\t&2 && 3 && {9\over7} &\cr
& 3&3&\h&\h&1 && 3&1&\h&\t&2 && +1 && -{3\over14} &\cr
& \d          && 3&3&\h&\t&1 && +\sqrt{8/3} && -1/\sqrt{6} &\cr
& \d          && 3&3&\h&\t&2 && 0 && \sqrt{3/10} &\cr
& 3&1&\h&\t&2 && 3&1&\h&\t&2 && 3 && -{6\over5} &\cr
& \d          && 3&1&\t&\h&2 && +1 && {1\over5} &\cr
& 3&3&\h&\t&1 && 3&3&\h&\t&1 && -5 && -1 &\cr
& \d          && 3&3&\t&\h&2 && +1 && {1\over5}&\cr
& 3&3&\h&\t&2 && 3&3&\h&\t&2 && 3 && 14\sqrt{2}/9 &\cr
& \d          && 3&3&\t&\h&2 && 1 && -7\sqrt{2}/27 &\cr
& 4&3&\h&\h&1 && 4&3&\h&\t&2 && 0 && -\sqrt{5/6} &\cr
& \d          && 4&3&\h&\t&1 && +\sqrt{8/3} && 1/(3\sqrt{6}) &\cr
& 4&5&\h&\h&1 && 4&3&\h&\t&2 && 0 && -\sqrt{2/3} &\cr
& \d          && 4&3&\h&\t&1 && 0 && -{1\over3}\sqrt{10/3} &\cr
& 4&3&\h&\t&2 && 4&3&\h&\t&2 && 3 && 3 &\cr
& \d          && 4&3&\t&\h&1 && +1 && -{1\over2} &\cr
& 4&3&\h&\t&1 && 4&3&\h&\t&1 && -5 && {1\over3} &\cr
& \d          && 4&3&\t&\h&1 && +1/3 && {5\over18} &\cr
& 4&4&\h&\h&0 && 4&2&\h&\t&2 && 3 && -{12\over7} &\cr
& 4&2&\h&\t&2 && 4&2&\h&\t&2 && 3 && -{12\over7} &\cr
\vrule height12pt depth9pt width0pt
& \d          && 4&2&\t&\h&2 && +1 && {2\over7} &\cr
}\hrule}$$
\vskip.4in\vfill\eject
}
\endinsert

The last published versions of the coupled-channel $R$-matrix model for this
system were the ``new $S$'' model of Ref.~[3] and the ``old $S$''
model.$^{2,\,3}$ Those versions were the first to include the CBM pole in the
$f$-matrix with the associated larger value $r_0 = 1.05$~fm.  The ``new $S$''
version takes isobar width into account.  It was consistent with the
experimental value of $\sigma\left(pp\to pp\pi^+\pi^-\right)$ at $T_L =
800$~MeV.  Its only important discrepancy from experiment was that it
predicted too much inelasticity for $600\,\hbox{MeV}<T_L\le 800\,\hbox{MeV}$.
It predicted $\eta(800\,\hbox{MeV}) = 0.84$ while the phase shift analysis
(PSA) prefers 0.98.  The excess of inelasticity was a result of the strength
of coupling to the $N\!\Delta$ channel, which in turn was required to produce
the correct energy dependence of the real phase $\delta$.  Switching coupling
strength to the high threshold $\Delta\Delta$ or $N\!N^*$ channels reduced the
inelasticity, but produced excessive two-pion production as in the ``old $S$''
model.

The OPE potential in the diagonal $N\!\Delta$ interaction, now included, is
repulsive, decreasing the inelasticity near the $N\!\Delta$ threshold.
Because of the importance of two-pion production we also include the OPE
transition potentials between the $N\!\Delta$ and $\Delta\Delta$ states.  We
can now transfer $N\!N-N\!\Delta$ coupling strength to $N\!N-\Delta\Delta$
coupling, but the increased production of these two-pion channels is
transferred back to one-pion channels by the $N\!\Delta-\Delta\Delta$
coupling.

The $\Delta\Delta\left({}^5D_0\right)$ channel is now included along with the
$\Delta\Delta\left({}^1S_0\right)$ channel, because of the former's strong OPE
coupling.  But the width of this channel is neglected, because its orbital
angular momentum barrier means that its effective threshold is higher than its
mass threshold.  On the other hand, the $N\!N^*\left({}^1S_0\right)$ channel
has been dropped because of its weak OPE coupling to the $N\!N$ channel.  It
would be difficult to distinguish its effect from that of the
$\Delta\Delta\left({}^1S_0\right)$ channel.  The result is a good fit to both
$\delta$ and $\eta$ over the whole energy range.  Simultaneously the
production of the two-pion producing $\Delta\Delta$ channel is kept low enough
so that the contribution to $\sigma\left( pp\to pp\pi^+\pi^-\right)$ is about
one-third of the experimental total value$^{10}$ of 3 $\mu b$ for $T_L =
800$~MeV.

The potential matrix elements involving isobar channels are (Eq.~(2.2))
$$\eqalign{V_T \left[ N\!N \left( {}^1S_0\right), N\!\Delta\left(
{}^5D_0\right) \right] &= 0.311 m_\pi v_2 \cr
V_T \left[ N\!N\left( {}^1S_0\right), \Delta\Delta\left( {}^1S_0\right) \right]
&= 0.172 m_\pi v_0 \cr
V_T \left[ N\!N \left( {}^1 S_0\right), \Delta\Delta \left( {}^5 D_0\right)
\right] &= -0.1744 m_\pi v_2 \cr
V_T\left[ N\!\Delta \left({}^5 D_0\right), \Delta\Delta \left( {}^1S_0\right)
\right] &= -0.0696 m_\pi v_2 \cr
V_T \left[ N\!\Delta \left( {}^5 D_0\right),\Delta\Delta\left( {}^5 D_0\right)
\right] &= 0.139 m_\pi v_0 -0.0696 m_\pi v_2\qquad \hbox{and}\cr
V_D \left[ N\!\Delta \left( {}^5 D_0\right) \right]
&= -0.0383 v_0 + 0.1935 v_2\ \ .
\cr}$$

We note that the direct and exchange contributions to $V_D$ (Figs.~3a and 3b)
cancel in the coefficient of $v_0$, but add in the coefficient of $v_2$
producing a substantial repulsive potential.

The pole of the $0^+$, $\left[ q\left(1S_{1/2}\right)\right]^6$ configuration
is at $W_p = 2.71$~GeV/$c^2$, corresponding to $T_L = 2.04$~GeV.  Only the
$\Delta\Delta\left( {}^1S_0\right)$ and $N\!N\left( {}^1S_0\right)$ channels
overlap with the configuration, and they have residues (Eq.~(2.12))
$$\eqalign{
\rho\sub{N\!N,N\!N} &= 0.149\,\hbox{GeV}\ \ ,\cr
\rho\sub{N\!N,\Delta\Delta0} &= 0.133\,\hbox{GeV}\qquad \hbox{and}\cr
\rho\sub{\Delta\Delta0,\Delta\Delta0} &= 0.118718\,\hbox{GeV}\ \ .\cr}$$
Here and below we use the total spin $S$ of the channel in the subscripts to
distinguish otherwise ambiguous channels.

The constant, non-vanishing, elements of the $f$-matrix are adjusted to fit
the phase parameters and the singlet scattering length $a^{pp}_s =
-7.82\,\hbox{fm}$.  They are $f^{\,0}_{N\!N,N\!N} = 83.5743$,
$f^{\,0}_{N\!\Delta,N\!\Delta} = 0.0$, $f^{\,0}_{\Delta\Delta 0,
\Delta\Delta 0} = 3.0$, $f^{\,0}_{\Delta\Delta 2,\Delta\Delta 2} = 5.0$,
$f^{\,0}_{N\!N,N\!\Delta} = 6.0$ and $f^{\,0}_{N\!N,\Delta\Delta2} = - 25.2$.
The
resulting phases are compared with the PSA results$^{17,\,18}$ in Fig.~2,
showing a very good fit.  The PSA results of Refs.~[19,20] generally agree
with those of Refs.~[17,18] but Ref.~[19] does not go to the lower energies,
and Ref.~[20] has oscillations not present in the other PSA results or in
ours.

\goodbreak\bigskip
\noindent{\bf B.\quad The \eb{$N\!N\left({}^3P_0\right)$}\ System}
\medskip\nobreak
In the last published version$^8$ the $N\!N\left( {}^3P_0\right)$ channel was
coupled to the $N\!\Delta\left( {}^3P_0\right)$ channel only.  We now also
include the $\Delta\Delta\left( {}^3P_0\right)$ but neglect the effect of its
width.  In Ref.~[8] the phase shift fit was improved for $T_L<300$~MeV by
reducing the pion coupling to $g^2_{N\!N\pi}/4\pi=13.0$.  However, with the
increased value of $r_0$ used now (1.05~fm instead of 0.75~fm) we fit well
with $g^2_{N\!N\pi}/4\pi = 14.4$ as in all the other partial waves.  The added
diagonal OPE potential for $N\!\Delta$ is repulsive.

The potential matrix elements are
$$\eqalign{V_T\left[ N\!N\left( {}^3P_0\right),N\!\Delta\left(
{}^3P_0\right)\right] &= 0.2075 m_\pi v_0 + 0.1037m_\pi v_2 \cr
V_T \left[ N\!N\left({}^3P_0\right),\Delta\Delta\left( {}^3P_0\right)\right]
&= 0.130 m_\pi v_0 - 0.052 m_\pi v_2\qquad\hbox{and}\cr
V_D \left[ N\!\Delta\left( {}^3P_0\right)\right] &= 0.1418 m_\pi v_0 + 0.0135
m_\pi v_2 \ \ .\cr}$$

As seen in Fig.~3, a very good fit to the PSA is obtained with the non-zero
$f$-matrix elements
$f^{\,0}_{N\!N,N\!N} = 11.7$, $f^{\,0}_{N\!\Delta,N\!\Delta} = 2.0$,
$f^{\,0}_{\Delta\Delta,\Delta\Delta}=2.0$, $f^{\,0}_{N\!N,N\!\Delta} = 1.2$,
and $f^{\,0}_{N\!N,\Delta\Delta} = -6.5$.

We note that, both in our model results and the most recent PSA, the value of
$\eta$ monotonically decreases below 1~GeV, unlike those of Ref.~[8].  This is
in part due to the repulsive diagonal OPE potential in $N\!\Delta$.  The
result is that the previous indication of a resonance in the Argand plot (a
counter-clockwise motion) is now gone for $T_L<1$~GeV.

\goodbreak\bigskip
\noindent{\bf C.\quad The \eb{$N\!N\left( {}^3P_1\right)$}\ System}
\medskip\nobreak
Reference [1] contains the last published version of the
$N\!N\left({}^3P_1\right)$ system.  It uses the smaller $r_0$ and was coupled
to a single $N\!\Delta(P)$ channel.  The width of the $N\!\Delta$ channel
could then only be taken into account when $V_T$ was ignored.  We now couple
to the $N\!\Delta\left({}^3P_1\right)$, $N\!\Delta\left({}^5P_1\right)$,
$N\!\Delta\left({}^5F_1\right)$ and the $\Delta\Delta\left( {}^3P_1\right)$
channels.  The width is taken into account only for the $N\!\Delta\left(
{}^5P_1\right)$ channel, which is strongly core coupled as well as having a
low angular momentum barrier and a low mass threshold.  The other additions
and changes common to all partial waves have also been made.  The resulting
potential matrix elements are
$$\eqalignno{V_T\left[ N\!N\left( {}^3P_1\right), N\!\Delta\left(
{}^3P_1\right)\right] &= 0.2075m_\pi v_0 - 0.05186m_\pi v_2 + 0.8 {e^{-2m_\pi
r} \over r} \cr
V_T\left[ N\!N \left( {}^3P_1\right), N\!\Delta\left( {}^5P_1\right)\right]
&= - 0.2087 m_\pi v_2 + 3.0 {e^{-2m_\pi r}\over r} \cr
V_T \left[ N\!N\left( {}^3P_1\right), N\!\Delta\left( {}^5 F_1\right)\right]
&= 0.1704 m_\pi v_2 - 2.5 {e^{-2m_\pi r}\over r} \cr
V_T \left[ N\!N
\left( {}^3P_1\right),\Delta\Delta \left( {}^3 P_1\right)\right]
&= 0.13 m_\pi v_0 + 0.026 m_\pi v_2 \cr
V_D \left[ N\!\Delta\left( {}^3P_1\right)\right] &= 0.1418 m_\pi v_0
- 0.00675 m_\pi v_2 \cr
V_D \left[ N\!\Delta\left( {}^5P_1\right) \right] &= -0.1163 m_\pi v_0
+ 0.0809 m_\pi v_2\cr
V_D\left[ N\!\Delta\left({}^5F_1\right)\right] &= -0.1163 m_\pi v_0
+ 0.924 m_\pi v_2\ \ .\cr}$$
In this case, fitting the PSA required the addition of the phenomenological
two-pion range transition potentials to the OPE components connecting $N\!N$
to $N\!\Delta$ channels.  These phenomenological components substitute for
$\rho$-exchange tails as well as actual two-pion exchange.  They usually, as
here, decrease the coupling strength of OPE.  Although their numerical
coefficients are large, because of the difference in the $r$-dependent form,
they are dominated by the OPE tensor terms.

The fitted non-zero $f$-matrix elements are
(as before the numerals 1 and 2 in the subscripts denote the spin for $P$
states, while the 3 denotes the $F$ state)
$f^{\,0}_{N\!N,N\!N} = 17.915$,
$f^{\,0}_{N\!\Delta 1,N\!\Delta 1} = 2.0$,
$f^{\,0}_{N\!\Delta 2,N\!\Delta 2} = 10.0$,
$f^{\,0}_{N\!\Delta 3,N\!\Delta 3} = 4.0$,
$f^{\,0}_{\Delta\Delta,\Delta\Delta} = 5.0$,
$f^{\,0}_{N\!N,N\!\Delta 2} = -4.0$,
and $f^{\,0}_{N\!N, \Delta\Delta} = 10.85$.
As shown in Fig.~4, the fits to PSA are very good.

\vfill\eject
\noindent{\bf D.\quad The \eb{$N\!N\left( {}^3P_2 - {}^3F_2\right)$}\ System}
\medskip
The last published version of the $N\!N\left( {}^3P_2 - {}^3 F_2\right)$
system$^8$ already includes the effect of the width of the isobars.  In
addition to the general changes made to OPE couplings and the diagonal
$N\!\Delta$ potential, the $N\!\Delta$ ${}^3P_2$, ${}^5P_2$, ${}^3F_2$ and
${}^5F_2$ channels are each now treated independently, without combining the
two $P$ states and the two $F$ states.  The effect of width in the two $F$
states is neglected because of the angular momentum barrier and the vanishing
$f$-matrix coupling to the $N\!N$ states.  The fit to the PSA is
better than in Ref.~[8] and no phenomenological two-pion range potentials are
now needed.  The potentials involving isobars are
$$\eqalign{V_T\left[ N\!N\left({}^3P_2\right), N\!\Delta\left(
{}^3P_2\right)\right] &= 0.2075v_0 + 0.01037 v_2\cr
V_T\left[ N\!N\left( {}^3F_2\right), N\!\Delta \left( {}^3P_2\right)\right]
&= -0.0762v_2\cr
V_T\left[ N\!N\left( {}^3P_2\right), N\!\Delta \left( {}^5P_2\right)\right]
&= -0.0933 v_2 \cr
V_T\left[N\!N\left({}^3F_2\right),N\!\Delta \left({}^5P_2\right)\right]
&= 0.0762
v_2 \cr
V_T\left[ N\!N\left( {}^3P_2\right), N\!\Delta\left( {}^3F_2\right)\right]
&= -0.0762 v_2 \cr
V_T\left[ N\!N\left( {}^3F_2\right), N\!\Delta \left( {}^3 F_2\right)\right]
&= 0.2075 v_0 + 0.0415 v_2 \cr
V_T\left[ N\!N\left( {}^3P_2\right), N\!\Delta\left( {}^5 F_2\right)\right]
&= 0.1867 v_2\cr
V_T\left[ N\!N\left( {}^3F_2\right), N\!\Delta \left( {}^5 F_2\right)\right]
&= 0.1525 v_2\cr
V_D \left[ N\!\Delta\left( {}^3P_2\right)\right] &= 0.1418v_0 + 0.0014v_2\cr
V_D\left[N\!\Delta\left( {}^5P_2\right)\right] &= - 0.1163 u_0 - 0.0809v_2\cr
V_D\left[N\!\Delta\left({}^3F_2\right)\right] &= 0.1418v_0 + 0.0054v_2\cr
V_D\left[N\!\Delta\left( {}^5F_2\right)\right]
&= -0.1163 v_0 + 0.0231 4v_2 \cr}$$
The $f$-matrix choice to fit the PSA is, for the non-vanishing elements,
$f^{\,0}_{N\!N\!P,N\!N\!P} = 13.917$,
$f^{\,0}_{N\!N\!F,N\!N\!F} = 12.0$,
$f^{\,0}_{N\!\Delta P1, N\!\Delta P1} = 5.5$,
$f^{\,0}_{N\!\Delta P2, N\!\Delta P2} = 2.5$,
$f^{\,0}_{N\!\Delta F1, N\!\Delta F1} = 4.0$,
$f^{\,0}_{N\!\Delta F2, N\!\Delta F2} = 4.0$,
$f^{\,0}_{N\!N\!P,N\!N\!F}=-2.8$,
$f^{\,0}_{N\!N\!P,N\!\Delta P1}=10.30$,
and $f^{\,0}_{N\!N\!P,N\!\Delta P2}=-1.2$.

The fit to the PSA shown in Fig.~5 is a substantial improvement to that of
Ref.~[8].  For two phase parameters it is not as good a fit as we achieve in
most of the partial waves, but is adequate in most respects.  There is little
significance to the large difference between our $\phi\sub{2}$ and that of
Arndt {\it et al.\/}$^{18}$ That parameter is very sensitive to the data and
is strongly constrained in Ref.~[18] by their assumption of only one coupled
isobar channel.

\goodbreak\bigskip
\noindent{\bf E.\quad The \eb{$N\!N\left({}^1D_2\right)$}\ System}
\medskip\nobreak

In Ref.~[3] the $f$-matrix pole arising from the $2^+$ state of the
$\left[q\left( {}^1S_{1/2}\right)\right]^6$ graph configuration was included.
The configuration energy, as given by the CBM quark dynamics and
$r_0=1.05$~fm, is $W_p=2.880$~GeV.  The configuration contains components of
the $N\!\Delta\left({}^5S_2\right)$ and $\Delta\Delta\left({}^5S_2\right)$
systems, and the non-vanishing residue-matrix components are (Eq.~(2.12))
$$
\eqalign{\rho\sub{N\!\Delta2,N\!\Delta 2} &= 0.260\ \hbox{GeV}\cr
\rho\sub{N\!\Delta2,\Delta\Delta2} &= 0.116276\ \hbox{GeV}\qquad\hbox{and}\cr
\rho\sub{\Delta\Delta2,\Delta\Delta2} &= 0.052\ \hbox{GeV}\ \ .\cr}
$$
The rest of this previous model was described in Ref.~[8], where the width of
the $\Delta$ is taken into account.  The OPE coupling to the
$\Delta\Delta\left({}^5S_2\right)$ channel was neglected then.

Because its vanishing angular momentum barrier compensates for the higher
threshold, we now include the $\Delta\Delta\left({}^5S_2\right)$ channel as
well as the $N\!\Delta\left({}^5D_2\right)$ and $N\!\Delta\left({}^5S_2\right)$
channels included in Ref.~[8].  The width of the $\Delta\Delta$ channel is
neglected.  Among the general coupling changes made, the most important here
is the inclusion of OPE diagonal potentials in the $N\!\Delta$ channels.  They
are attractive, greatly improving the shape of the inelasticity parameter, as
will be shown below.  The required potentials are
$$\eqalign{V_T\left[ N\!N\left({}^1D_2\right),
N\!\Delta\left({}^5S_2\right)\right] &= 0.1392 v_2\cr
V_T\left[ N\!N\left({}^1D_2\right), N\!\Delta\left( {}^5 D_2\right)\right] &= -
0.1663 v_2\cr
V_T\left[ N\!N\left({}^1D_2\right), \Delta\Delta\left({}^5S_2\right)\right] &=
-0.0794v_2\cr
V_D\left[ N\!\Delta\left( {}^5S_2\right)\right]
&= -0.0383 v_0\quad \hbox{and}\cr
V_D \left[ N\!\Delta\left( {}^5D_2\right)\right]
&= -0.0383v_0-0.0415v_2~~.\cr}$$
There are no phenomenological two-pion range potentials.  The adjusted
$f$-matrix elements are
$f^{\,0}_{N\!N,N\!N}=15.60$,
$f^{\,0}_{N\!\Delta S,N\!\Delta S}=1.0$,
$f^{\,0}_{N\!\Delta D, N\!\Delta D} = 10.0$,
$f^{\,0}_{\Delta\Delta,\Delta\Delta}=1.0$,
$f^{\,0}_{N\!\Delta,N\!\Delta S} =0.6$,
and $f^{\,0}_{N\!N,\Delta\Delta}=-6.50$.

The resulting fit, shown in Fig.~6, is quite good for $\delta$.  For $\eta$,
in contrast to the fit of Ref.~[8], the rapid drop to $T_L=650$~MeV followed
by a sudden leveling is reproduced.  The remaining defect is that the model
value of $\eta$ in the 700~MeV region is too small by about 0.06 and the model
value increases beyond 800~MeV.  It is the attractive OPE diagonal $N\!\Delta$
potential which increases the effect of coupling at the lower energies, but is
less effective at the higher energies, as required.

\goodbreak\bigskip
\noindent{\bf F.\quad The \eb{$N\!N\left({}^3F_3\right)$}\ System}
\medskip\nobreak
The Ref.~[8] version of the $N\!N\left({}^3F_3\right)$ system, the last
previously published, has the opposite deficiency to that just encountered in
the ${}^1D_2$ state with respect to the inelasticity.  The experimental $\eta$
keeps decreasing uniformly for $515~\hbox{MeV}<T_L <800~\hbox{MeV}$, while the
model $\eta$ begins to level off near 650~MeV.  The OPE diagonal interaction
in the $N\!\Delta\left({}^5P_3\right)$ channel is dominantly repulsive (see
below).  The rapid drop of $\eta$ above threshold can be maintained by
increasing the coupling to the $N\!N$ channel, and the drop is then maintained
to higher energy as the long-range repulsive potential becomes less
effective.

We now also include the
$N\!\Delta\left({}^3F_3\right)$
and
$N\!\Delta\left({}^5F_3\right)$
channels.
The width of all $N\!\Delta$ channels is considered.  The potentials are
$$
\eqalign{
V_T \left[ N\!N \left( {}^3 F_3 \right),
N\!\Delta \left({}^5 P_3 \right)\right]
&= -0.1822 v_2\cr
V_T \left[ N\!N \left( {}^3 F_3 \right),
N\!\Delta \left({}^3 F_3 \right)\right]
&= 0.2075 v_0 - 0.0520 v_2\cr
V_T \left[ N\!N \left( {}^3 F_3 \right),
N\!\Delta \left({}^5 F_3 \right)\right]
&= 0.0696 v_2\cr
V_D\left[ N\!\Delta\left( {}^5P_3\right)\right]
&= -0.1163 v_0 + 0.0231 v_2\cr
V_D\left[ N\!\Delta\left( {}^3F_3\right)\right]
&= 0.1418v_0 - 0.0067v_2 \quad\hbox{and}\cr
V_D\left[N\!\Delta\left({}^5F_3\right)\right]
&=-0.1163v_0-0.0424v_2\ \ .\cr}
$$
No phenomenological two-pion potentials are required.  The non-zero fitted
$f$-matrix parameters are $f^{\,0}_{N\!N,N\!N}=5.90$,
$f^{\,0}_{N\!\Delta2,N\!\Delta2}=0.8$, $f^{\,0}_{N\!\Delta1,N\!\Delta1}=1.0$,
and
$f^{\,0}_{N\!N,N\!\Delta2}=-0.3$.  As shown in Fig.~7, a very good fit is now
obtained.

\goodbreak\bigskip
\noindent{\bf G.\quad The \eb{$N\!N\left({}^3F_4 - {}^3H_4\right)$}\ System}
\medskip\nobreak
In the case of the $N\!N\left({}^3F_4-{}^3H_4\right)$ system the last
published version dates back to the no-isobar results of Ref.~[1].  In
Ref.~[3] it was remarked that coupling to isobars would not have a large
effect for $T_L<1$~GeV because of the large angular momentum barrier ($L\ge
3$).  In fact, the change due to the couplings described below is not large,
but there is a substantial improvement to the fit to $\delta\left({}^3
F_4\right)$.

The couplings to both the $N\!\Delta\left({}^3F_4\right)$ and
$N\!\Delta\left({}^5F_4\right)$ channels are included,
and the width of the $\Delta$ is not neglected in either channel.
The potential matrix elements are
$$\eqalign{
V_T \left[ N\!N\left( {}^3F_4\right), N\!\Delta\left({}^5 F_4\right)\right] &=
-0.116v_2\cr
V_T\left[ N\!N\left( {}^3H_4\right), N\!\Delta\left( {}^5 F_4\right)\right] &=
0.1032 v_2 - 2.1\, e^{-m_\pi r}/r \cr
V_T \left[ N\!N\left( {}^3F_4\right), N\!\Delta\left( {}^3 F_4\right)\right] &=
0.207 v_0 + 0.017 v_2 \cr
V_T \left[ N\!N \left( {}^3 H_4\right), N\!\Delta\left( {}^3F_4\right)\right]
&=
-0.078v_2 + 1.6\, e^{-m_\pi r}/r \cr
V_D \left[ N\!\Delta \left( {}^5 F_4\right)\right] &= - 0.116v_0 - 0.0578v_2
\quad \hbox{and}\cr
V_D\left[ N\!\Delta \left({}^3F_4\right)\right] &= 0.142v_0 + 0.0025v_2\ \
.\cr}$$

The strongly attractive OPE diagonal potential in the
$N\!\Delta\left({}^5F_4\right)$ state compensates substantially for the effect
of the orbital angular momentum barrier, to the extent that too much
low-energy coupling occurs.  Consequently, a best fit requires the inclusion
of a phenomenological two-pion range contribution in the
$N\!N\left({}^3H_4\right)$ coupling to $N\!\Delta\left({}^5F_4\right)$ and
$N\!\Delta\left({}^3F_4\right)$, as above.

The fitted non-zero $f$-matrix constants are
$f^{\,0}_{N\!N\!F,N\!N\!F} = 8.49$,
$f^{\,0}_{N\!N\!H, N\!N\!H} = 30.0$,
$f^{\,0}_{N\!\Delta 2, N\!\Delta 2} = 5.0$,
$f^{\,0}_{N\!\Delta 1, N\!\Delta 1} = 5.0$,
$f^{\,0}_{N\!N\!F, N\!N\!H} = -2.0$,
$f^{\,0}_{N\!N\!F, N\!\Delta 2} = -3.90$,
and $f^{\,0}_{N\!N\!F, N\!\Delta 1} = 4.80$.
The results, shown in Fig.~8, are in good agreement with the PSA except for
$\delta\left({}^3H_4\right)$.  The predicted values of
$\delta\left({}^3H_4\right)$ are too attractive over most of the energy
range.  The coupling and the boundary condition have little effect because of
the $L=5$ barrier. If the PSA results are correct this appears to represent a
deficiency of the potential in the $N\!N$ sector for large $L$, perhaps
indicating a repulsive quadratic spin-orbit term.

\goodbreak\bigskip
\noindent{\bf H.\quad The \eb{$N\!N\left({}^1G_4\right)$}\ System}
\medskip\nobreak

As in the previous $N\!N\left( {}^3F_4-{}^3H_4\right)$ case, modified versions
of the $N\!N\left( {}^1G_4\right)$ system have not been published since the
uncoupled result of Ref.~[1].  In this case the coupling is potentially more
important because it couples to the $L=2$ $N\!\Delta\left({}^5D_4\right)$
channel.  The width of that channel is included.  The isobar related potential
matrix elements are
$$
\eqalign{
V_T\left[N\!N\left({}^1G_4\right), N\!\Delta\left({}^5D_4\right)\right]
&= 0.1663v_2 - 1.80\, e^{-m_\pi r}/r
\quad\hbox{and}\cr
V_D \left[ N\!\Delta\left( {}^5 D_4\right)\right]&= - 0.0383v_0 + 0.0553 v_2
\ \ .\cr}$$
A phenomenological two-pion range contribution in the coupling potential is
again required to cancel a coupling effect which is too strong at medium-range.
The $f$-matrix constants are
$f^{\,0}_{N\!N,N\!N} = 9.80$,
$f^{\,0}_{N\!\Delta,N\!\Delta}=2.5$
and $f^{\,0}_{N\!N,N\!\Delta} = 4.0$.
As shown in Fig.~9 a very good fit to the PSA
is obtained.

\goodbreak\bigskip
\noindent{\bfmone IV. \enskip Amplitudes and Observables}
\medskip\noindent{\bf A. \enskip The Method}\medskip
With respect to the observables, several conventions and notations are in use
in the literature and it is sometimes difficult to determine their
relationship.  This is only partly due to notation in which zero-spin labels
were omitted and triple-scattering experiments were not considered or clearly
distinguished.  It mainly owes its origin to the different definitions of the
coordinate systems, both in the center of mass (c.m.) and in the laboratory.
Here, we stick to the Madison convention (positive normal along $\vec{k}_i
\times \vec{k}_f$) and the Saclay convention${}^{21}$ (same normal used in
both frames but also in the three different sets of basis vectors in the
laboratory) and avoid the Basle convention which uses two normals.  Five
common choices of orthonormal basis vectors are compared in Table II.

\goodbreak\midinsert
\centerline{\bfmone Table II}
\medskip
\centerline{\hbox{\vbox{\offinterlineskip
\hrule height 0.75pt
\halign{
#{\vrule height 15pt depth 10pt width 0.75pt}&
\quad#\hfil\quad&\vrule#&
\hfil\enskip$#$\enskip\hfil&\vrule#&
\hfil\enskip$#$\enskip\hfil&\vrule#&
\hfil\enskip$#$\enskip\hfil&\vrule#&
\hfil\enskip$#$\enskip\hfil&\vrule#&
\hfil\enskip$#$\enskip\hfil&\vrule#&
\hfil\enskip$#$\enskip\hfil&\vrule#& 
\hfil\enskip$#$\enskip\hfil&\vrule#&
\hfil\enskip$#$\enskip\hfil&\vrule#&
\hfil\enskip$#$\enskip\hfil&\vrule#&
\hfil\enskip$#$\enskip\hfil&\vrule#&
\hfil\enskip$#$\enskip\hfil&\vrule#&
\hfil\enskip$#$\enskip\hfil&\vrule#&
\hfil\enskip$#$\enskip\hfil&#{\vrule width 0.75pt}\cr 
&{\bf Group}&&\multispan{11} \hfil {\bf Center of mass system} \hfil &&
\multispan{13} \hfil {\bf Laboratory system} \hfil &\cr
\noalign{\hrule}
&Saclay${}^{21}$&&k_i&&k_f&&n&&\ell&&m&&0&&k&&s&&k'&&s'&&k''&&s''&&n&\cr
\noalign{\hrule}
&Wolfenstein${}^{22}$&&p&&p'&&n&&P&&K&&%
${\ninerm omitted}$&&k&&{\ }&&k'&&s&&k'_t&&s_t&&n&\cr
\noalign{\hrule}
&Halzen-Thomas${}^{23}$&&z&&{\ }&&y&&\dag&&\ddag&&
${\ninerm omitted}$&&\ell&&s&&\ell&&s&&\ell&&s&&n&\cr
\noalign{\hrule}
&Argonne${}^{24}$&&\multispan{11} \hfil same as above \hfil
&&L&&S&&L&&S&&L&&S&&N&\cr
\noalign{\hrule}
&Raynal${}^{25}$&&k_i&&k_f&&n&&P&&K&&%
${\ninerm omitted}$&&z&&-x&&{~}&&{~}&&{~}&&{~}&&y&\cr
\noalign{\hrule height 0.75pt}}}}}
{\par\baselineskip=14pt\leftskip=0.5in\rightskip=0.5in
{\tenrm A selection of some representative sets of basis vectors encountered
in the literature both in the center-of-mass and laboratory systems.  The
vectors listed in a column are all equal.  One notes that ($\dag$)
${\ell = x \sin{\theta\over2} + z \cos{\theta\over2}}$ and ($\ddag$)
${m = x \cos{\theta\over2} - z \sin{\theta\over2}}$~.}\par}
\endinsert

The notation (Beam, Target; Scattered, Recoil) to express observables is
adopted by Argonne whereas Saclay used (Scattered, Recoil; Beam, Target).  The
position of a given spin label in such a quadruplet thus specifies to which
particle the frame is attached.  Taking into account this information, the use
of superscripts in the lab system by Saclay, in order to distinguish the
incident (no superscript), scattered (prime) and recoil (double dash) particle
is unnecessary and is being progressively abandoned in observable names.  It
follows from Table II that $C_{\!K\!K} \equiv C_{mm00} = A_{00mm}$,
$C_{\!P\!P} \equiv C_{\ell\ell00} = A_{00\ell\ell}$ and $C_{\!K\!P} =
C_{\!P\!K} \equiv C_{\ell m00} = -A_{00\ell m}$ (in each case the last
equality comes from time reversal invariance).  The Wolfenstein
system${}^{22}$ is then simply connected to the Saclay system.${}^{21}$

The theoretical group at Argonne is used to the Halzen-Thomas
notation.${}^{23}$~ Their vectors $(x,y,z)$ define the center-of-mass helicity
frame.  The experimental group at Argonne has capitalized the spin labels in
the laboratory system.  Raynal's notation${}^{25}$ defines the spin labels
$x$, $y$, $z$ in the laboratory system but these should not be confused with
the Argonne center-of-mass same labels $x$, $y$, $z$.  The polarization labels
$X$, $Y$, $Z$ used in the SAID system correspond to Raynal's index $x$, $y$,
$z$.  Convention also differs, as the sideways direction $\vec{x}$ is chosen
to be parallel to $-\vec{s}$.  Such a choice, however, is not consistent with
the Saclay right-handed frame attached to the incident particle in the lab
system and for which $\vec{n}=\vec{k}\times\vec{s}$.  This is the reason why
some minus signs appear when connecting Raynal's formalism${}^{25}$ to the
Saclay one, {\it e.g.\/} $A_{\!Z\!X} = - A_{00sk}$.  Still another different
convention but now in the center of mass can be found in the literature.  In
some older papers, the defined vector $\vec{m}$ has its sign opposite to the
one of the Saclay formalism.  This was made so that in the non-relativistic
case the c.m.  vector $\vec{m}$ coincides in direction with the lab momentum
$\vec{k}''$ of the recoil particle, which is not the case in the Saclay
formalism.  Again connecting identical observables defined with respect to two
different triplets of basis vectors in the same coordinate system involves a
few minus signs (as many as there are in index).  A final source of confusion
is that the Wolfenstein transfer parameters $R'_t$ and $A'_t$ are defined with
respect to $-\vec{k}''$, which changes the sign of the normal in the lab frame
attached to the recoil particle.  The connection with the Saclay formalism is
then $R'_t=-K_{0k''s0}$ and $A'_t=-K_{ok''k0}$.

\goodbreak\midinsert
\centerline{\bfmone Table III}\medskip
\centerline{\hbox{\vbox{\offinterlineskip
\hrule height 0.75pt\halign{
#{\vrule height 15.5pt depth 8.5pt width 0pt}{\vrule width0.75pt}&
\hfil\enskip$#$\enskip\hfil&\vrule#&
\hfil\enskip$#$\enskip\hfil&\vrule#&
\hfil\enskip$#$\enskip\hfil&\vrule#&
\hfil\enskip$#$\enskip\hfil&#{\vrule width 0.75pt}\cr 
& $SAID${}^{18} && $Wolfenstein${}^{22} &&%
$Saclay${}^{21} && $Argonne${}^{24} &\cr
\noalign{\hrule}
& DSG && I_0 && \sigma~{\hbox{or}}~{d\sigma\over d\Omega} && \sigma &\cr
& P && P && P_{n000} = A_{00n0} = A_{000n} && P &\cr
& D && D && D_{n0n0} = D_{0n0n} && D_{\!N\!N} &\cr
& DT && D_t && K_{n00n} = K_{0nn0} && K_{\!N\!N} &\cr
& R && R && D_{s'0s0} = D_{0s''0s} && D_{SS} &\cr
& RP && R' && D_{k'0s0} = D_{0k''0s} && D_{SL} &\cr
& A && A && D_{s'0k0} = D_{0s''0k} && D_{LS} &\cr
& AP && A' && D_{k'0k0} = D_{0k''0k} && D_{LL} &\cr
& AYY && C_{nn} && C_{nn00} = A_{00nn} && C_{\!N\!N} &\cr
& AZZ &&$\lower12pt\hbox{\vbox{%
\hbox{$C_{\!P\!P} \cos^2{\theta\over2} + C_{\!K\!K} \sin^2{\theta\over2}$}
\hbox{$\quad \phantom{1} + C_{\!K\!P}\sin\theta$}}}$%
&& A_{00kk} && C_{LL} &\cr
\omit{\vrule height7pt width 0.75pt}&&&&&&&&\cr
& AXX &&$\lower12pt\hbox{\vbox{%
\hbox{$C_{\!P\!P} \sin^2{\theta\over2} + C_{\!K\!K} \cos^2{\theta\over2}$}
\hbox{$\quad \phantom{1} - C_{\!K\!P}\sin\theta$}}}$%
&& A_{00ss} && C_{SS} &\cr
\omit{\vrule height6pt width 0.75pt}&&&&&&&&\cr
& AZX &&$\lower12pt\hbox{\vbox{%
\hbox{${1\over2} (C_{\!K\!K} - C_{\!P\!P}) \sin\theta$}
\hbox{\quad $\phantom{1} + C_{\!K\!P}\cos\theta$}}}$%
&& -A_{00sk} = -A_{00ks} && -C_{SL} &\cr
\omit{\vrule height5pt width 0.75pt}&&&&&&&&\cr
}\hrule height 0.75pt}}}
{\par\baselineskip=14pt\leftskip=1in\rightskip=1in
\noindent
{\tenrm Name and sign conventions for the observables computed in the present
paper in relation to four different formalisms often encountered in the
literature.}\par\medskip}
\endinsert

Keeping these five choices in mind, we compare the observables that we will
compute, in Table III.  One can see that the names used in the SAID system
come from different groups.

We calculate the observables from the bilinear combinations of the helicity
amplitudes, which themselves are constructed from the partial-wave
$S$-matrix.  The spin partial-wave expansions of the $s$-channel helicity
amplitudes follow:
$$
\eqalignno{
\phi_1 &= {1\over 2ik} \sum_{J=0}^\infty
\left\{ (2J+1) \alpha\sub{J,0} + (J+1) \alpha\sub{J,+} + J \alpha\sub{J,-}
+ 2 \sqrt{J(J+1)} \alpha^J \right\}  P_J (\cos\theta)
&{(4.1a)}\cr
\phi_2 &= {1\over 2ik} \sum_{J=0}^\infty
\left\{ -(2J+1) \alpha\sub{J,0} + (J+1) \alpha\sub{J,+} + J \alpha\sub{J,-}
+ 2 \sqrt{J(J+1)} \alpha^J \right\}  P_J (\cos\theta)
&{(4.1b)}\cr
\phi_3 &= {1\over 2ik} \sum_{J=0}^\infty
\left\{ (2J+1) \alpha\sub{J,1} + J \alpha\sub{J,+} + (J+1) \alpha\sub{J,-}
- 2 \sqrt{J(J+1)} \alpha^J \right\} \times &{(4.1c)}\cr
& \hskip 3.5in \times
\left[ P_J (\cos\theta) + {1-\cos\theta \over J(J+1)} P'_J(\sin\theta) \right]
\cr
\phi_4 &= {1\over 2ik} \sum_{J=0}^\infty
\left\{ -(2J+1) \alpha\sub{J,1} + J \alpha\sub{J,+} + (J+1) \alpha\sub{J,-}
- 2 \sqrt{J(J+1)} \alpha^J \right\} \times &{(4.1d)}\cr
& \hskip 3.5in \times
\left[ -P_J (\cos\theta) + {1+\cos\theta \over J(J+1)} P'_J(\sin\theta) \right]
\cr
\phi_5 &= {1\over 2ik} \sum_{J=0}^\infty
\left\{ \sqrt{J(J+1)}
\left( \alpha\sub{J,+} - \alpha\sub{J,-} \right)
- \alpha^J \right\}
\ {\sin\theta \, P'_J (\sin\theta) \over \sqrt{J(J+1)}}
&{(4.1e)}\cr}
$$
where $P_J(\cos\theta)$ are the Legendre polynomials of order $J$ (total
angular momentum), $P'_J(\sin\theta)$ are derivatives of $P_J(\cos\theta)$
with respect to their argument, and $k$ is the c.m.~relative momentum.  The
energy-dependent $\alpha$ parameters describe the different possible $S$- and
$L$- coupled states: $\alpha\sub{J,0}$ represents the $S$ singlet,
$\alpha\sub{J,1}$ represents the uncoupled triplet, and $\alpha\sub{J,\pm}$,
$\alpha^{J}$ represent the $L$ coupled triplet states.  It is understood in
the above sums that, when we encounter a $J=0$ pole, then the corresponding
$\alpha$ parameter has to be set equal to zero.  Consequently,
$\alpha\sub{0,1} = \alpha\sub{0,+} = \alpha\sub{0,-} = \alpha^{0} = 0$ and the
${}^3\!P_0$ lacks a coupled partner.  The $\alpha\sub{J,\pm}$ are the diagonal
partial-wave amplitudes for tensor coupled states while the $\alpha^{J}$ are
the off-diagonal ones.

If $\delta$ is the real part of the complex phase shift and $\sqrt{\eta}$ is
the exponential of the imaginary part, then our inelasticity parameter $\eta$
is related to Bugg's $\rho$ [Ref.~17] by the equality $\eta = \cos^2 \rho$.
Thus $\eta=1$ in the elastic region $T_L \lessim 300$ MeV.  For the
parametrization of the partial-wave amplitudes in terms of the phase shift
parameters, that definition leads to
$$\eqalignno{
\alpha\sub{J,0} &= \eta\sub{J,0} \exp (2i\delta_{J,0}) - 1
&{(4.2a)}\cr
\alpha\sub{J,1} &= \eta\sub{J,1} \exp (2i\delta_{J,1}) - 1
&{(4.2b)}\cr
\alpha\sub{J,\pm} &=
\eta\sub{J,\pm} \cos 2 \bar{\epsilon}\sub{J} \exp (2i\delta_{J,\pm}) -1
&{(4.2c)}\cr
\alpha^J &= i \sqrt{\eta\sub{J,-} \eta\sub{J,+}} \sin 2\bar{\epsilon}\sub{J}
\exp \left[ i \left( \delta_{J,-} + \delta_{J,+} + \phi\sub{J} \right) \right]
&{(4.2d)}\cr}$$
where $\eta\sub{J,\pm}=1$ in the elastic region and $\bar{\epsilon}\sub{J}$ is
the elastic (or ``real'') mixing angle between $J,+$ and $J,-$ states.  The
inelastic (or ``imaginary'') mixing angle $\phi\sub{J}$ is non zero only above
the inelastic threshold and its numerical value remains small just above it.
However, for states of small angular momentum barrier ({\it e.g.\/}
$\phi\sub{2}$), its value rises rapidly with energy and has a point of
inflection at the $\Delta$ resonance threshold.  The mixing angles
$\phi\sub{2}$ and $\phi\sub{4}$ presented in relation (4.2d) should not be
confused with the helicity amplitudes of eqs. (4.1b) and (4.1d).  In that
representation, the $\bar{\epsilon}$'s, $\delta$'s and $\phi$'s are real
numbers and $0\leq\eta\leq 1$.  In our figures for $\eta$'s and $\delta$'s, we
replace the set of label subscripts by the spectroscopic notation
(${}^{2S+1}\!L_J$).  The coupled channel parameters obey a sub-unitarity
inequality${}^{26}$ because of their other coupling to isobar channels.
Finally, the ${}^3\!P_0$ wave has the same representation as the singlet
states.

The observables can be computed from the c.m.~invariant amplitudes used by
Saclay (formulae can be found in Ref.~[21]) and related to the helicity
amplitudes by
$$\eqalignno{
a &= {1\over2}
\left[ \left( \phi_1 + \phi_2 + \phi_3 - \phi_4 \right) \cos\theta
- 4 \phi_5 \sin\theta \right] &{(4.3a)}\cr
b &= {1\over2}
\left( \phi_1 - \phi_2 + \phi_3 + \phi_4 \right) &{(4.3b)}\cr
c &= {1\over2} \left( -\phi_1 + \phi_2 + \phi_3 + \phi_4 \right) &{(4.3c)}\cr
d &= {1\over2} \left( \phi_1 + \phi_2 - \phi_3 + \phi_4 \right) &{(4.3d)}\cr
e &= -{i\over2}
\left[ \left( \phi_1 + \phi_2 + \phi_3 - \phi_4 \right) \sin\theta
+ 4 \phi_5 \cos\theta \right] &{(4.3e)}\cr}$$
This particular set of amplitudes is transformed into several other ones also
used in the literature, in Refs.~[21] and~[27].  The normalization is such
that
$${d \sigma \over d \Omega} = {1\over2}
\left\{ |a|^2 + |b|^2 + |c|^2 + |d|^2 + |e|^2 \right\}
\eqno{(4.4)}$$
is in $mb/sr$.  For $J\leq4$, the partial waves used are the coupled channel
ones of Section III, that is
${}^1\!S_0$,
${}^3\!P_0$,
${}^3\!P_1$,
${}^3\!P_2 - {}^3\!F_2$,
${}^1\!D_2$,
${}^3\!F_3$,
${}^3\!F_4 - {}^3\!H_4$ and
${}^1\!G_4$.
In fitting the energy dependence of the total cross sections, a compromise is
reached in using the waves ${}^3\!H_5$ and ${}^1\!I_6$ given by the
Feshbach-Lomon model without isobar coupling.${}^{4}$~ The same $N\!N$
coupling constant in the $\pi$- and $2\pi$- sector (14.40) is used
throughout.  The unitarized OPE phases begin with ${}^3\!H_6 - {}^3\!J_6$,
${}^3\!J_7$, $\ldots$ until $J=12$, the starting value for the OPE Born
amplitude.  The Coulomb interaction is included, except for the total cross
sections obtained by the optical theorem.  In the latter case, the pure
Coulomb amplitudes and phases are removed but the electro-nuclear
interferences in the model phases are kept.

The above prescriptions for treating OPE and Coulomb effects in our model are
of course slightly different than those of Bugg${}^{17}$ or Arndt.${}^{18}$\
Nevertheless, we will see that the different treatments lead to minor
differences in the polarization observables at low energy (142 MeV) and that
the differences become appreciable only when $\theta \lessim 5^{\circ}$ and
only at higher energies.  The observables of Table III are calculated at
energies of 142, 210, 325, 400, 475, 515, 580, 615, 685, 750, 800 and 970
MeV.  The excitation functions to be displayed below cover the whole range of
energies, $T_L \leq 1$ GeV, whereas angular distributions are presented at
142, 515 and 800 MeV only (except for $A_{zz}$ for which we add 580 and 685
MeV).

\bigskip\goodbreak
\line{\bf B. \enskip The Results \hfil}\medskip\nobreak
The excitation curves for the total unpolarized ($\sigma\sub{TOT}$) and
polarized ($\Delta\sigma\sub{L}$ and $\Delta\sigma\sub{T}$) cross sections are
shown in Figure 10.  The experimental points are in general well reproduced by
the model predictions as the latter follow the correct curvatures and
structures in the energy dependence, especially for $\Delta\sigma\sub{L}$.
However $\Delta\sigma\sub{T}$ does deviate from the data in that the position
of the model peak is about 50 MeV (laboratory energy) higher that the data
peak, and the prediction falls off more slowly that the data after the peak.
This is correlated with the deviation, already noted in Section III, of
$\eta({}^1\!D_2)$ from the PSA result for $T_L > 600$ MeV.  The model of $\eta
({}^1\!D_2)$ continues to decrease after the PSA result levels off at 600 MeV,
and it does not rise to the PSA value until $T_L \geq 800$ MeV.  Because
$\Delta\sigma\sub{T}$ increases with decreasing $\eta({}^1\!D_2)$, the model
peak is shifted to higher energy and the value is higher than the experimental
value until $T_L \approx 900$ MeV.

The excitation functions at $90^{\circ}$ c.m.~for ${d\sigma\over d\Omega}$,
$R$, $R'$, $A$, $A'$, $D$, $C_{nn00}$, $A_{00kk}$, $A_{00ss}$ and $C_{\!K\!P}$
are compared with the data points between $89^\circ$ and $91^\circ$ compiled
in the SAID system,${}^{18}$ and are displayed in Figure 11.  Our predictions
for $C_{KK}$ are also compared with Arndt PSA results.  For identical
particles at $\theta_{cm} = {\pi\over2}$, other theoretical predictions are
$P=0,\ A_{00sk}=0$ and $D_{n0n0} = K_{0nn0}$ (the few existing $D_t$
experimental points are thus plotted together within the $D$ experimental
set).  We note that $R'$ and $A'$ are not independent quantities at this
angle.  For $pp$ scattering at $90^\circ$, one has ${A\over A'} = -{R'\over R}
= \tan \theta_1$ with $\theta_1$ being the lab scattering angle.  Theory
implies that the polarization-transfer parameters here are not independent
from the depolarization and rotation parameters since $R_t(90^\circ) =
R(90^\circ)$, $R'_t(90^\circ)=R'(90^\circ)$, $A_t(90^\circ)=-A(90^\circ)$ and
$A'_t(90^\circ)=-A'(90^\circ)$.  Consequently we could have plotted any
independent measurements of $R_t$, $R'_t$, $A_t$, $A'_t$ together with the
usual Wolfenstein set, as we did for $D_t$ and $D$.  However, at this time no
data points exist for the transfer parameters in the range $88^\circ < \theta
< 92^\circ$ up to 1 GeV.  Under the same circumstances, one also has $A_{00kk}
+ A_{00ss} = C_{nn00} - 1$ because of the Pauli principle.  Other constraints
at $90^\circ$ c.m.~on the limits of some spin observables from amplitude
analysis can be found in Ref.~[28].

The fit of the $90^\circ$ excitation functions to the data is generally very
good.  The only substantial deviation from data for $T_L \leq 800$ MeV is for
${d\sigma\over d\Omega} (90^\circ)$ at 650~MeV~$<~T_L~\leq~800$~MeV.  In this
range the newest data is higher than the older data and our prediction.  Above
800 MeV, up to 1 GeV, our predictions for $D$, $A_{yy}$ and $A_{zz}$ decrease
with energy while the data remains nearly constant.

For the $R'$, $A'$, $A_{xx}$ and $C_{KK}$ excitation functions we also show
the result of the SAID PSA${}^{18}$ because the data is sparse or
non-existent.  The trend is similar to our prediction, with somewhat less
structure.

The comparison with the angular distributions is shown in Fig.~12.  The model
energies chosen as representative are $T_L = 142$, 515 and 800 MeV for
${d\sigma\over d\Omega}$, $P$, $R$, $R'$, $A$, $A'$, $D$, $D_t$, $A_{yy}$,
$A_{zz}$, $A_{zx}$, and $A_{xx}$.  In addition we compare with recent $A_{zz}$
data (LAMPF) at 580 and 685 MeV.  Again the Arndt PSA prediction${}^{18}$ is
also shown where the data is sparse or non-existent.  The experimental
energies compared with these curves are listed in the caption.  They are
mostly within 5 MeV of the model energy, but sometimes differ by as much as 10
MeV.  There also exists some Wolfenstein spin-transfer $R_t$ and $A_t$
data,${}^{18}$ but only at 800 MeV.  There are eight points of moderate to
poor accuracy.  None of the points are near $\theta_{cm} = 90^\circ$.  We have
not calculated the model predictions for $R_t$ and $A_t$ or any of the
triple-index spin observables.

The fit to the angular distributions is good, and has no important deviations
from the data for $T_L < 800$ MeV.  Most of the observables are also well
reproduced at $T_L = 800$ MeV.  However at this energy the model shows more
angular dependence than the data for the observables $P$, $R'$, $A$, $A_{yy}$,
and $A_{zz}$.  This trend persists at 580 and 685 MeV for $A_{zz}$, but the
statistical significance is small.  The problem may be associated with the
model value of $\delta ({}^3\!H_4)$.  This has the only significant deviation
from the PSA of any of the $\delta$'s.  The high $L$ value implies a strong
angular dependence, and our $\delta({}^3\!H_4)$ is significantly larger than
the PSA for $T_L > 500$ MeV.

\goodbreak\bigskip
\line{\bfmone V. \enskip Conclusions \hfil}\medskip\nobreak
We have presented the results of a model for $pp$ scattering that includes:

\item{(a)}
Coupled $N\!N$, $N\!\Delta$ and $\Delta\Delta$ channels for $J \leq 4$. The
effect of the width of the $\Delta$ is taken into account.

\item{(b)}
The $\pi$, $2\pi$, $\eta$, $\rho$ and $\omega$ meson exchange potentials in
the $N\!N$ sector, and $\pi$-exchange potentials connecting the isobar channel
sector to itself and to the $N\!N$ sector.  The coupling parameters used were
determined by experiments other than $N\!N$ scattering or from $SU(3)$ and
quark model relations.

\item{(c)}
At a separation radius $r_0$ an $R$-matrix boundary condition is imposed
representing valence quark degrees of freedom for $r < r_0$.  The energy
independent part of the $f$-matrix are parameters.  The poles are determined
by the energies of the quark configurations as given by the CBM.  Only the
poles of the lowest 6-quark configuration $(1s_{1\over2})^6$ are included,
affecting the ${}^1\!S_0$ and ${}^1\!D_2$ $pp$ channels.  The value of $r_0$
is determined by the physical requirements of the $R$-matrix method.

\smallskip
\noindent
The result is a very good fit to almost all the data for $T_L < 1$ GeV.  The
major exceptions are the higher energy of the peak in $\Delta\sigma\sub{T}$
and an excessive angular dependence in a few of the observables.  No other
model produces quality fits for $T_L > 300$ MeV and they especially fail in
describing the structures in $\Delta\sigma\sub{L}$ and $\Delta\sigma\sub{T}$.
The good degree to which these structures are fitted by this model strongly
implies that the structures are due to the effect of coupling to the
$N\!\Delta$ threshold in the $N\!N$~~${}^1\!D_2$ and ${}^3\!F_3$ channels.

The shift of the $\Delta\sigma\sub{T}$ peak is related to the fact that the
model $\eta({}^1\!D_2)$ structure is deeper than the PSA result.  The
inability of the model to more closely fit the PSA value of $\eta({}^1\!D_2)$
may be related to the omission in the model of coupling to the $\pi D$
channel.  This channel couples relatively strongly to the $N\!N ({}^1\!D_2)$.
Including it, in addition to the isobar channel coupling of the model, may
allow a simultaneous fit of the steep descent of $\eta({}^1\!D_2)$ just above
threshold and the weakened inelastic effect above 650 MeV laboratory energy.

The excess of angular structure in some of the observables for $T_L > 515$ MeV
is likely due to the model value of $\delta({}^3\!H_4)$ being larger than the
PSA value at the higher energies.  The difference is at most $0.5^\circ$, but
this can be significant because of the high statistical weight of the $J=4$
partial wave.  Because of the very large angular momentum barrier $(L=5)$ the
boundary condition parameter has negligible effect.  This indicates that the
medium range meson-exchange potential used here is inaccurate for large $L$.
A correction to our potential can arise from high order non-local effects
which give rise to quadratic spin-orbit type terms.  These were not included.

In spite of the above noted deficiencies the $pp$ data for $T_L < 1$ GeV is
overall well understood in terms of this model implying that most of the
physics has been included.  An important part of this physics is the
$R$-matrix boundary condition representing the quark degrees of freedom at
short range.  Even in those channels in which we ignore the pole terms
(because they are distant), the energy independent boundary condition acts
differently than any energy-independent non-singular potential.${}^{4}$~ In
fact it behaves as if the interior potential increases linearly with energy,
asymptotically becoming a hard core${}^{29}$, and the present results
reinforce that conclusion.

Of course the most interesting effects of the $R$-matrix core representation
come from the poles coinciding with quark configuration energies.  These give
rise to exotic quark resonances, exotic dibaryons in the present case, which
begin at $T_L > 1.8$ GeV.  The implications for structure in the observables
have already been given for the earlier forms of these model${}^{2,3}$ and
have interesting correlations with experiment.${}^{3,30,31}$~ Those
predictions will soon be updated with the present models, and the models
themselves will be extended to include the poles due to the $(1s_{1\over2})^5
(1p_{1\over2})$ configuration.  This modification to the odd-parity states
will negligibly affect the $T_L < 1$ GeV predictions.

The other obvious extension of this work is to update the $I=0$ partial wave
models and compare with the $np$ data.  That work is well underway.

\goodbreak\bigskip
\line{\bfmone Acknowledgements \hfil}
\medskip\nobreak

One of the authors (ELL) benefitted from the computer and other facilities of
P-division and LAMPF at the Los Alamos National Laboratory, and also those at
the Institut f\"ur Kernphysik, Forschungszentrum J\"ulich.  He is also
grateful for discussions on aspects of nucleon-nucleon experimental and PSA
results with Dr.~M.~McNaughton, and nucleon-nucleon theory with
Dr.~K.~Holinde.  Another author (PLF) is most grateful to John~Negele for the
warm hospitality he received during his visit to the MIT Center for
Theoretical Physics, where the paper was completed.  The other author (MA)
greatly appreciates the facilities provided by Professor M.~Banerjee and the
Nuclear Theory Grroup at the University of Maryland, which enabled him to
complete his research for this paper.

\vfill\eject

\centerline{\bf REFERENCES} {\par

\parskip=\medskipamount
\raggedright
\tolerance 10000

\item{(a)} Present address: Physics Department, University of Maryland,
College Park MD 20742.

\item{1.}
E. L. Lomon, {\it Phys. Rev.\/} {\bf D26}, 576 (1982).

\item{2.}
P. LaFrance and E. L. Lomon, {\it Phys. Rev.\/} {\bf D34}, 1341 (1986).

\item{3.}
P. Gonz\'alez, P. LaFrance and E. L. Lomon,
{\it Phys. Rev.\/} {\bf D35}, 2142 (1987).

\item{4.}
H. Feshbach and E. L. Lomon, {\it Ann. Phys.\/} (NY) {\bf 29}, 19 (1964).

\item{5.}
E. L. Lomon and H. Feshbach, {\it Ann. Phys.\/} (NY) {\bf 48}, 94 (1968).

\item{6.}
E. L. Lomon, {\it AIP Conference Proceedings\/} {\bf 110}, 117 (1984); {\it
Nucl. Phys.\/} {\bf A434}, 139c (1985); {\it J. Phys.\/} (Paris), Colloq. {\bf
46}, C2--329 (1985).

\item{7.}
P. LaFrance, {\it Can. J. Phys.\/} {\bf 67}, 1194 (1989).

\item{8.}
P. Gonz\'alez and E. L. Lomon, {\it Phys. Rev.\/} {\bf D34}, 1351 (1986).

\item{9.}
I. P. Auer {\it et al.\/}, {\it Phys. Rev. Lett.\/} {\bf 41}, 1436 (1978);
{\bf 41}, 354 (1978); K. Hidaka and A. Yokosawa, {\it Surv. High Energy
Phys.\/} {\bf 1}, 141 (1980).

\item{10.}
F. H. Cverna {\it et al.\/}, {\it Phys. Rev.\/} {\bf C23}, 1698 (1981).

\item{11.}
M. H. Partovi and E. L. Lomon, {\it Phys. Rev.\/} {\bf D2}, 1999 (1970).

\item{12.}
R. B. Wiringa, R. A. Smith and T. L. Ainsworth, {\it Phys. Rev.\/} {\bf C29},
1207 (1984).

\item{13.}
A de~Shalit and I. Talmi, {\it Nuclear Shell Theory\/} (Academic Press, 1963).

\item{14.}
A. Edmonds, {\it Angular Momentum in Quantum Mechanics\/}, second edition
(Princeton University Press, 1960).

\item{15.}
E. P. Wigner and I. Eisenbud, {\it Phys. Rev.\/} {\bf 72}, 29 (1947);
A. M. Lane and R. G. Thomas, {\it Rev. Mod. Phys.\/} {\bf 30}, 257 (1958).

\item{16.}
R. L. Jaffe and F. E. Low, {\it Phys. Rev.\/} {\bf D19}, 2105 (1979).

\item{17.}
D. V. Bugg, {\it Phys. Rev.\/} {\bf C41}, 2708 (1990).

\item{18.}

R. A. Arndt, J. S. Hyslop, III and L. D. Roper, {\it Phys. Rev.\/} {\bf D35},
128 (1987); the current SAID version, SP93, is used in the figures.

\item{19.}

J. Bystricky, C. Lechanoine-Leluc and F. Lehar, {\it J. Phys. France\/} {\bf
48}, 199 (1987).

\item{20.}
Y. Higuchi, N. Hoshizaki, H. Masuda and H. Nakao, {\it Prog. Theor. Phys\/}
{\bf 86}, 17 (1991).

\item{21.}
J. Bystricky, F. Lehar and P. Winternitz, {\it J.~Physique\/}~{\bf 39}, 1
(1978); P. LaFrance and P. Winternitz, {\it J.~Physique\/}~{\bf 41}, 1391
(1980).

\item{22.}
L. Wolfenstein, {\it Ann.~Rev.~Nucl.~Sci.\/}~{\bf 6}, 43 (1956).

\item{23.}
F. Halzen and G. H. Thomas, {\it Phys.~Rev.\/}~{\bf D10}, 344 (1974).

\item{24.}
A. Yokosawa, {\it Phys.~Rep.\/}~{\bf 64}, 49 (1980).

\item{25.}
J. Raynal, {\it Nucl.~Phys.\/}~{\bf 28}, 220 (1961).

\item{26.}
S. Klarsfeld, {\it Phys.~Lett.\/}~{\bf 126B}, 148 (1983).

\item{27.}
M. J. Moravcsik, J. Pauschenwein and G. R. Goldstein, {\it
J.~Phys.~France\/}~{\bf 50}, 1167 (1989).

\item{28.}
I. P. Auer {\it et al.\/}, {\it Phys.~Rev.\/}~{\bf D29}, 2435 (1984).

\item{29.}
E. L. Lomon, {\it Phys.~Rev.\/}~{\bf D12}, 3758 (1975).

\item{30.}
E. L. Lomon, {\it J.~Physique\/}~{\bf 58}, {\it Suppl.~Colloque\/}~{\bf C6},
363 (1990).

\item{31.}
E. L. Lomon, preprint CTP\#2207.
}\vfill\eject

\centerline{\bf FIGURE CAPTIONS}
\medskip

\def\d{\eb{$\diamondsuit$}}
\def\c{\raise.3ex\hbox{\eb{\ffex\char'015}}}
\def\x{\eb{$\times$}}
\def\s{$\square$}
\def\hb{\hfil\break}

\item{Fig.~1:}
Meson-exchange potential diagrams.  (a)--(c) Contributions to $N\!N$-$N\!N$
sector potentials; (d) for $N\!N$-$N$-isobar potentials; (e) for
$N\!N$-$\Delta\Delta$ potentials; (f) for $N\!\Delta$-$N\!\Delta$ potentials;
(g) for $N\!\Delta$-$\Delta N$ potentials.
\medskip

\item{Fig.~2:}
Phase parameters for $N\!N\left({}^1S_0\right)$ scattering.  Model -- solid
curve; PSA of Ref.~[18] -- dashed curve; PSA of Ref.~[17] -- \s.
\medskip

\item{Fig.~3:}
Phase parameters for $N\!N\left( {}^3P_0\right)$ scattering.
Notation as in Fig.~2.
\medskip

\item{Fig.~4:}
Phase parameters for $N\!N\left( {}^3P_1\right)$ scattering.
Notation as in Fig.~2.
\medskip

\item{Fig.~5:}
Phase parameters for the tensor coupled $N\!N\left( {}^3P_2-{}^3F_2\right)$
scattering.  The parameters determine the $S$-matrix $$\eqalign{S_{--} &=
\eta\left({}^3(J-1)_J\right)\cos 2\bar{\epsilon}_J\exp 2i\delta\left(
{}^3(J-1)_J\right)\ \ ,\cr S_{+-} &= S_{-+} =
i\left[\eta\left({}^3(J-1)_J\right) \eta\left( {}^3(J+1)_J\right)\right]^{1/2}
\sin 2\bar{\epsilon}_J \cr &\qquad \times\exp \left[ \delta\left(
{}^3(J-1)_J\right)+\delta\left({}^3(J+1)_J\right)+\phi\sub{J}\right]\ \ ,\cr
S_{++} &= \eta\left( {}^3(J+1)_J\right) \cos 2\bar{\epsilon}_J \exp
2i\delta\left( {}^3(J+1)_J\right)\ \ .\cr}$$ Notation as in Fig.~2.
\medskip

\item{Fig.~6:}
Phase parameters for $N\!N\left({}^1D_2\right)$ scattering.
Notation as in Fig.~2.
\medskip

\item{Fig.~7:}
Phase parameters for $N\!N\left({}^3F_3\right)$ scattering.
Notation as in Fig.~2.
\medskip

\item{Fig.~8:}
Phase parameters for the tensor coupled $N\!N\left({}^3F_4-{}^3H_4\right)$
scattering.  Parameters determine $S$-matrix as in Fig.~5.
Notations as in Fig.~2.
\medskip

\item{Fig.~9:}
Phase parameters for $N\!N\left({}^1G_4\right)$ scattering.
Notation as in Fig.~2.
\medskip

\item{Fig.~10:}
Excitation functions for the total unpolarized ($\sigma\sub{TOT}$)
and polarized ($\Delta\sigma\sub{L}$ and $\Delta\sigma\sub{T}$)
$pp$ cross sections for $T_L < 1$ GeV. \hb
\s: experimental values compiled in SAID.${}^{18}$ \hb
\d\ for $\Delta\sigma\sub{T}$: Saclay points (PE86). \hb
Solid curves: model.
\medskip\goodbreak

\item{Fig.~11:}
Excitation functions at $\theta_{cm} = 90^\circ$ for the differential cross
section ${d\sigma\over d\Omega}$, Wolfenstein parameters $R$, $R'$, $A$, $A'$,
$D$, second-rank asymmetry spin-tensor $A_{yy}$, $A_{zz}$, $A_{xx}$ ($A_{zx} =
0$ at this angle), and spin-correlations $C_{K\!P}$ and $C_{K\!K}$.  The two
last parameters are defined in the c.m. frame but the experimental points
shown do represent independent measurements.  Note that $C_{\!P\!P}(90^\circ)
= C_{\!K\!K}(90^\circ)$. \hb
\s: experimental values compiled in SAID. \hb
\x\ for ${d\sigma\over d\Omega}$: Saclay points (GA85). \hb
\x\ for $D$: independent $D_t$ points used since
$D(90^\circ) = D_t(90^\circ)$ (they are listed in SAID${}^{18}$). \hb Solid
curves: model. \hb Dashed curves: Arndt's PSA predictions.${}^{18}$
\medskip

\item{Fig.~12:}
Angular distributions computed at $T_L$ = 142, 515, and 800 MeV compared with
experimental values compiled in SAID$^{18}$ around each energy, for the
following observables.
\medskip
\halign{\hfil#&#\hfil\cr
${d\sigma\over d\Omega}$ at & 137 (\d), 144 (\c) and 144.1 (\x) MeV \cr
   & 513 (\s) and 516 (\x) MeV \cr
   & 788.7 (\d), 789 (\x), 795 (\s) and 800 (\c) MeV. \cr
$P$ at & 137 (\d), 140.7 (\s), 142 (\c) and 147 (\x) MeV \cr
   & 515 (\s), 515.3 (\d) and 517 (\x) MeV \cr
   & 790.1 (\d), 794 (\x), 796 (\s) and 800.9 (\c) MeV. \cr
$R$ at & 140 (\s) and 142 (\c) MeV \cr
   & 515.3 (\s) and 517 (\x) MeV \cr
   & 800 (\c,\s,\d) MeV. \cr
$R'$ at & 137.5 (\s) and 140.4 (\c) MeV \cr
   & 515.3 (\s) and 520 (\x) MeV \cr
   & 800 (\x,\s,\d) MeV. \cr
$A$ at & 139 (\s) and 143 (\c) MeV \cr
   & 517 (\s) MeV \cr
   & 800 (\c,\s,\d) MeV. \cr
$A'$ at & 800 (\x,\s,\d) MeV. \cr
$D$ at & 138 (\s), 142 (\c) and 143 (\x) MeV \cr
   & 515.3 (\s) and 517 (\x) MeV \cr
   & 800 (\c,\x,\s,\d) MeV. \cr
$D_t$ at & 517 (\s) MeV \cr
   & 800 (\c,\s) MeV. \cr
$A_{xx}$ at & 514 (\s) MeV \cr
   & 791 (\c) MeV. \cr
$-A_{zx}$ at & 514 (\s) MeV \cr
   & 793 (\x), 794 (\c) and 805.7 (\s) MeV. \cr
$A_{zz}$ at & 514 (\s) and 518.4 (\x) MeV \cr
   & 577 (\x), 583 (\c), 586.3 (\d) and 589 (\s) MeV \cr
   & 688 (\x) and 692 (\s) MeV \cr
   & 790.1 (\d), 793 (\x), 800 (\c) and 805.7 (\s) MeV. \cr
$A_{yy}$ at & 143 (\s) MeV \cr
   & 515 (\s) MeV \cr
   & 790.1 (\d), 794 (\x), 796 (\s) and 800 (\c) MeV. \cr}
Curves notated as in Fig.~11.
\par
\vfill
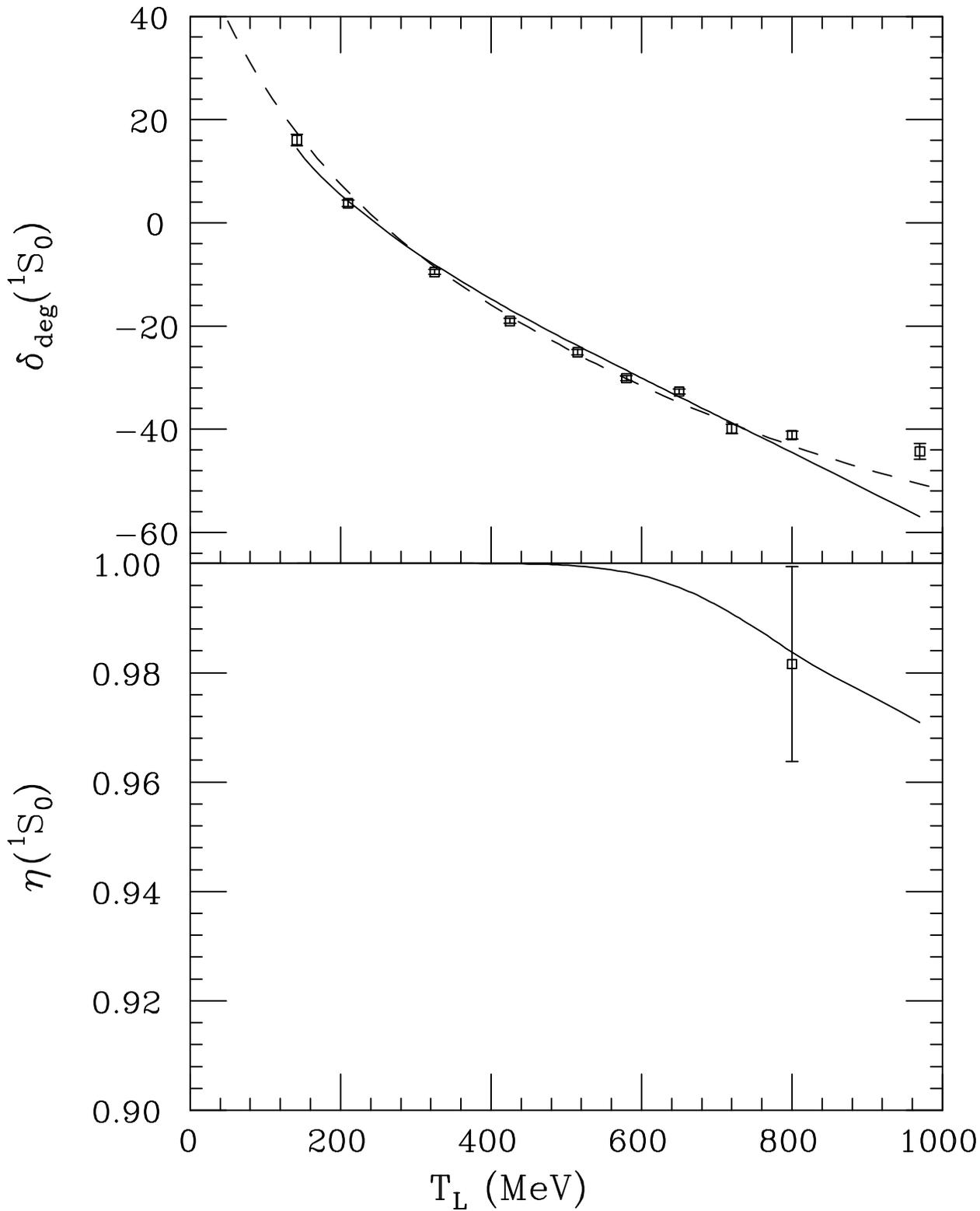
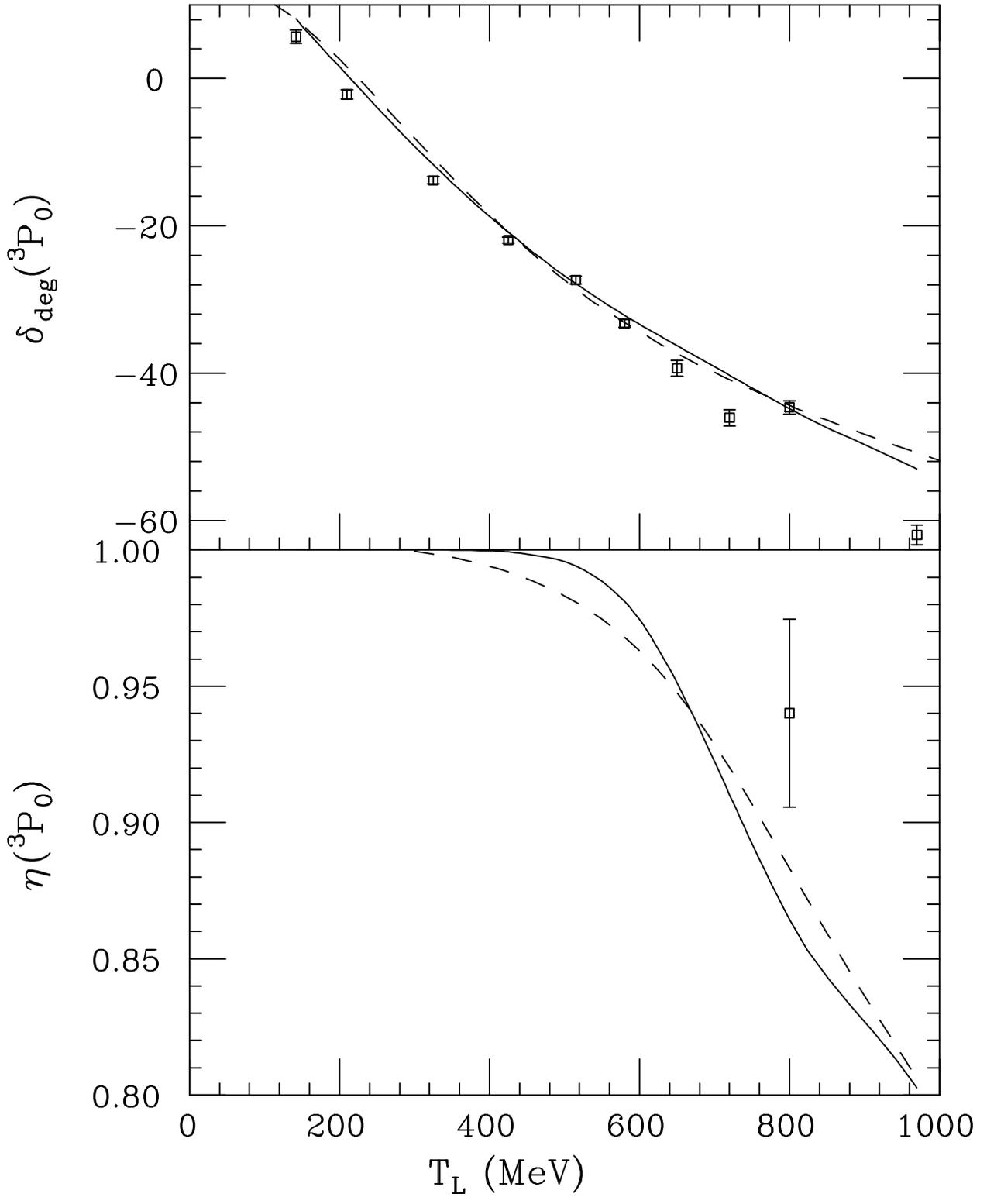
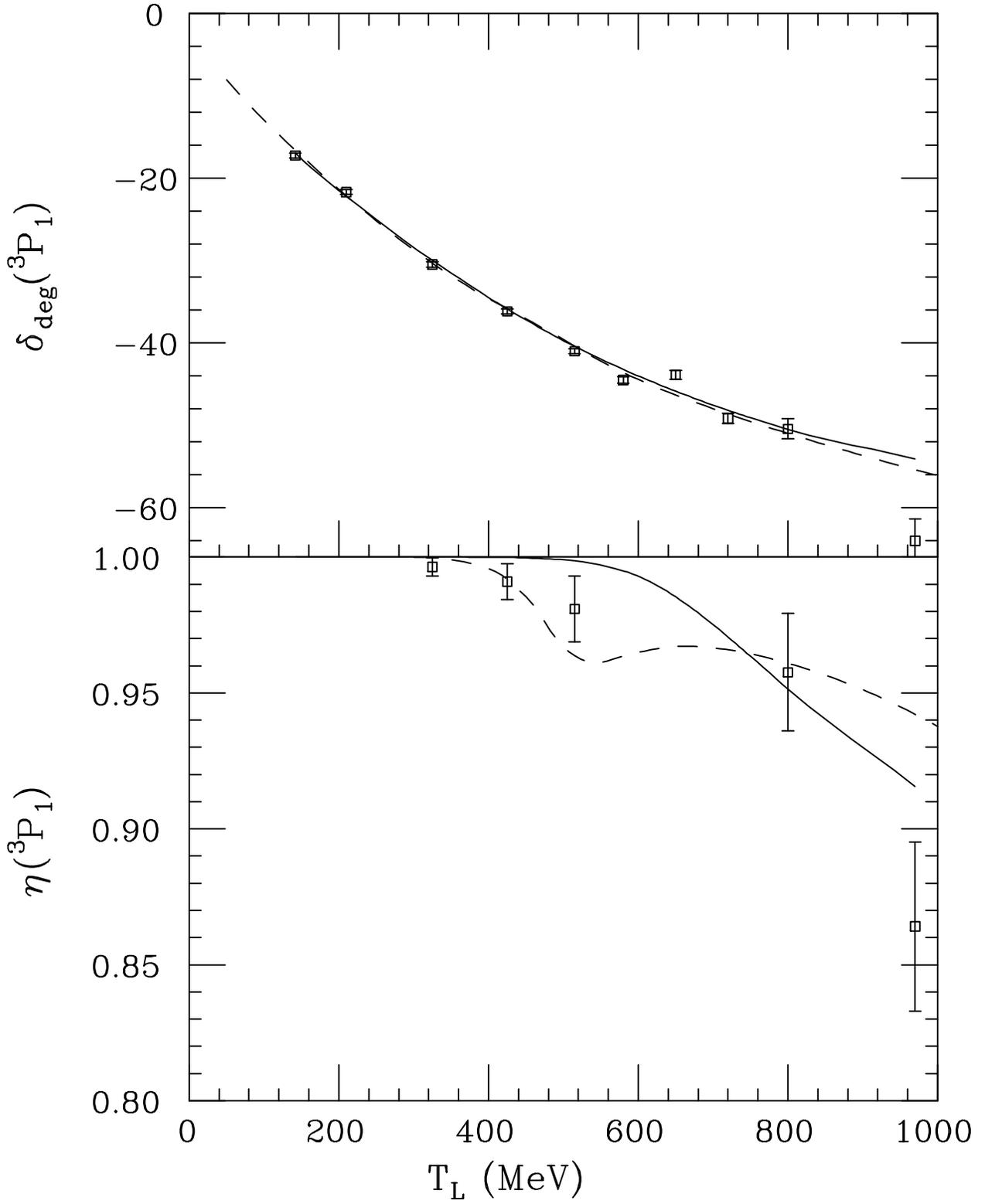
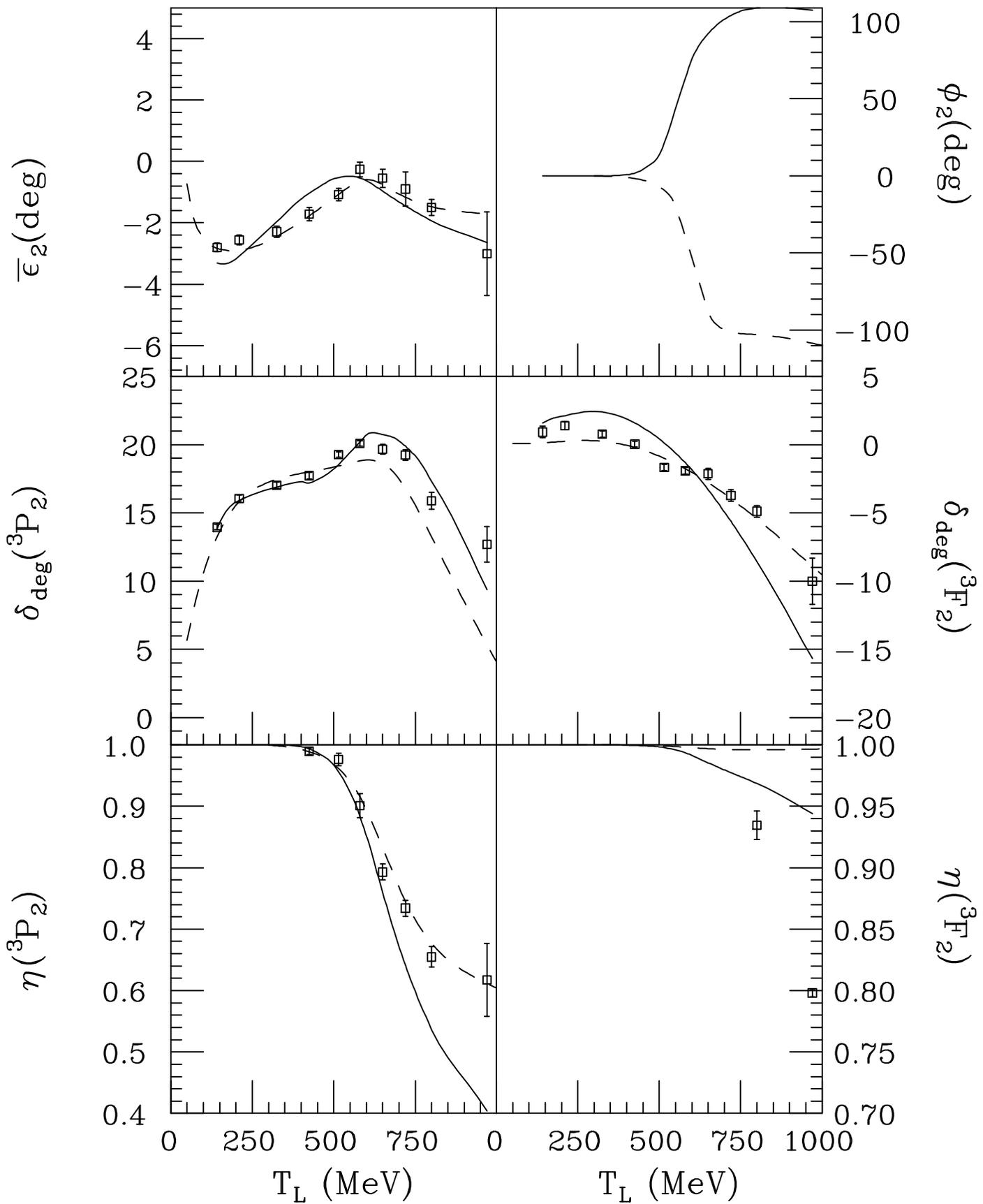
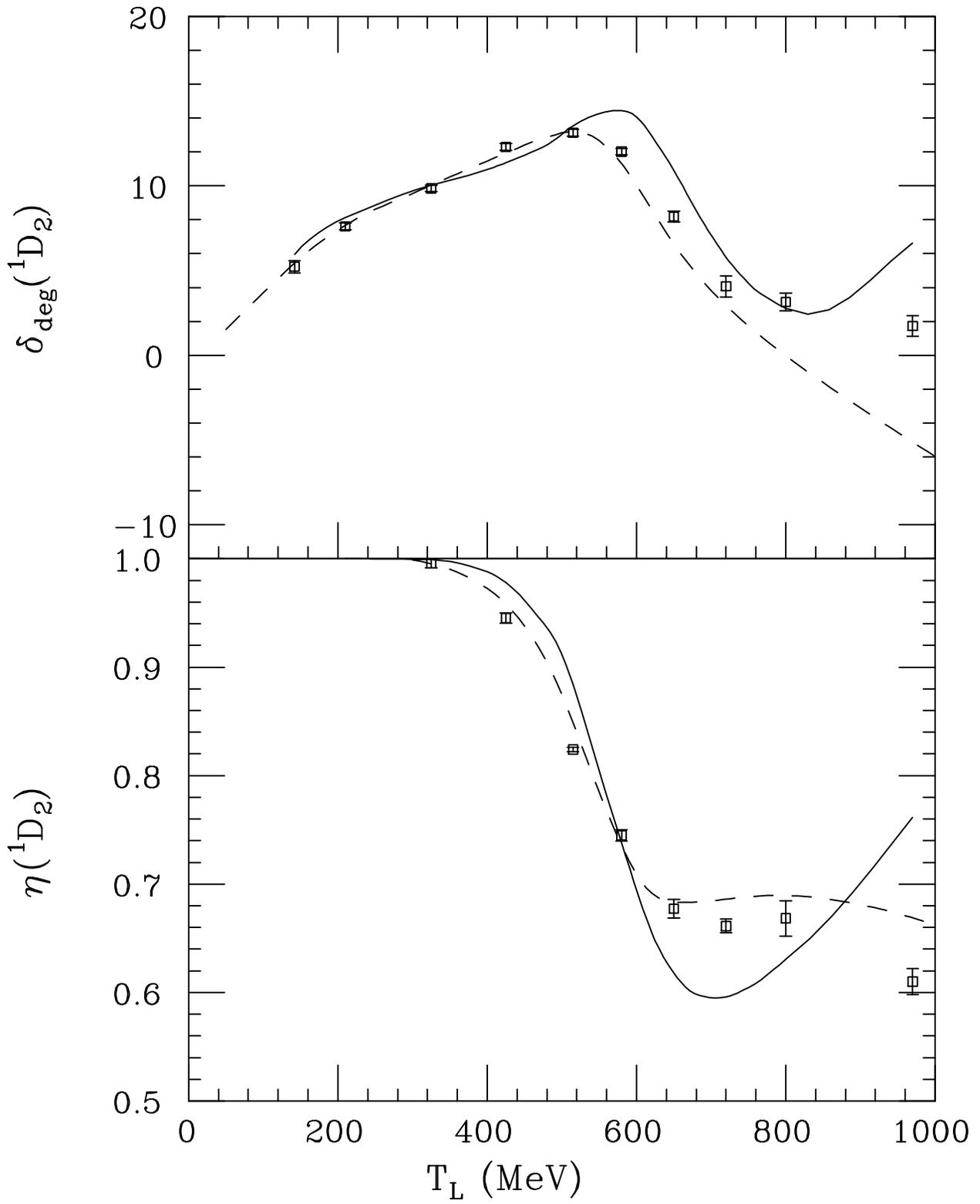
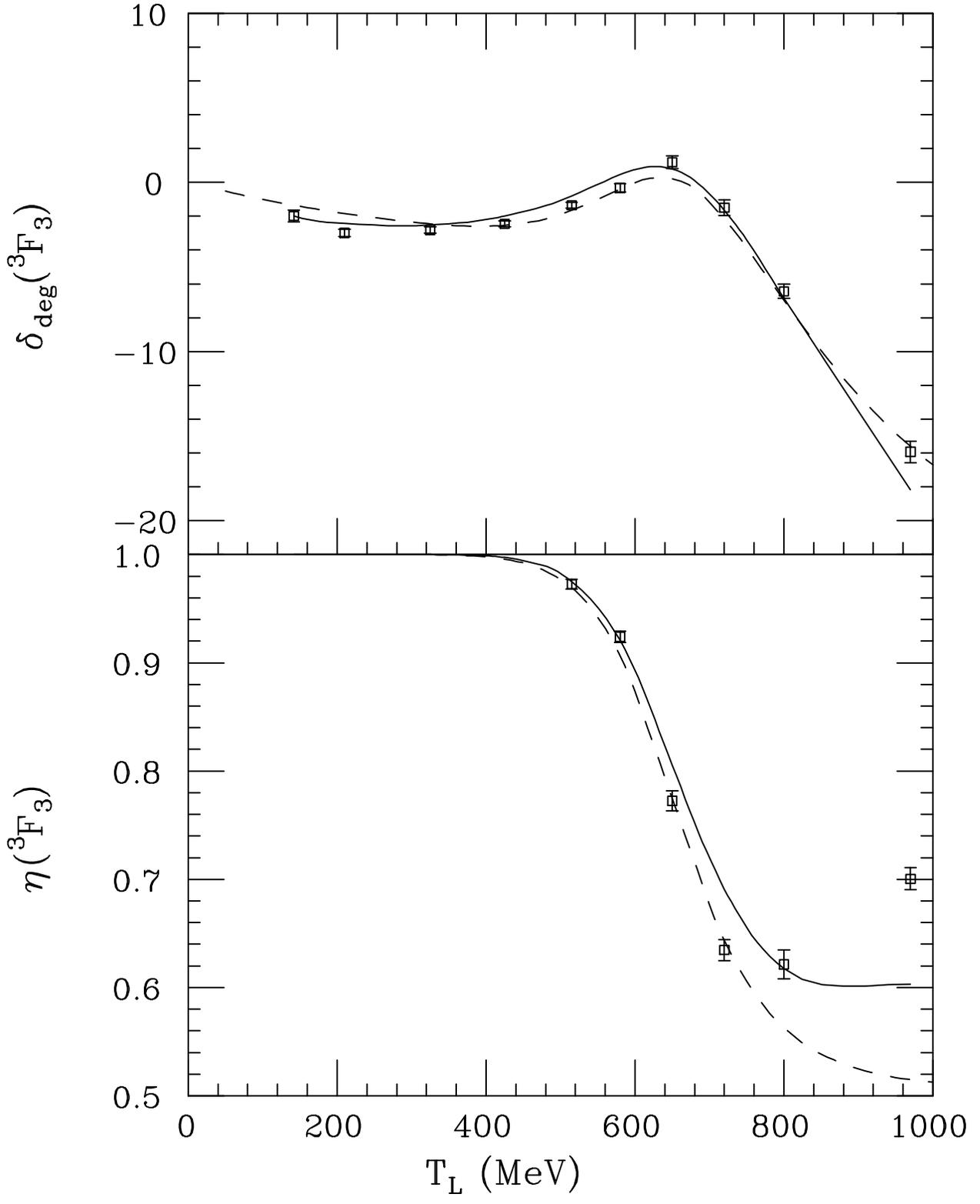
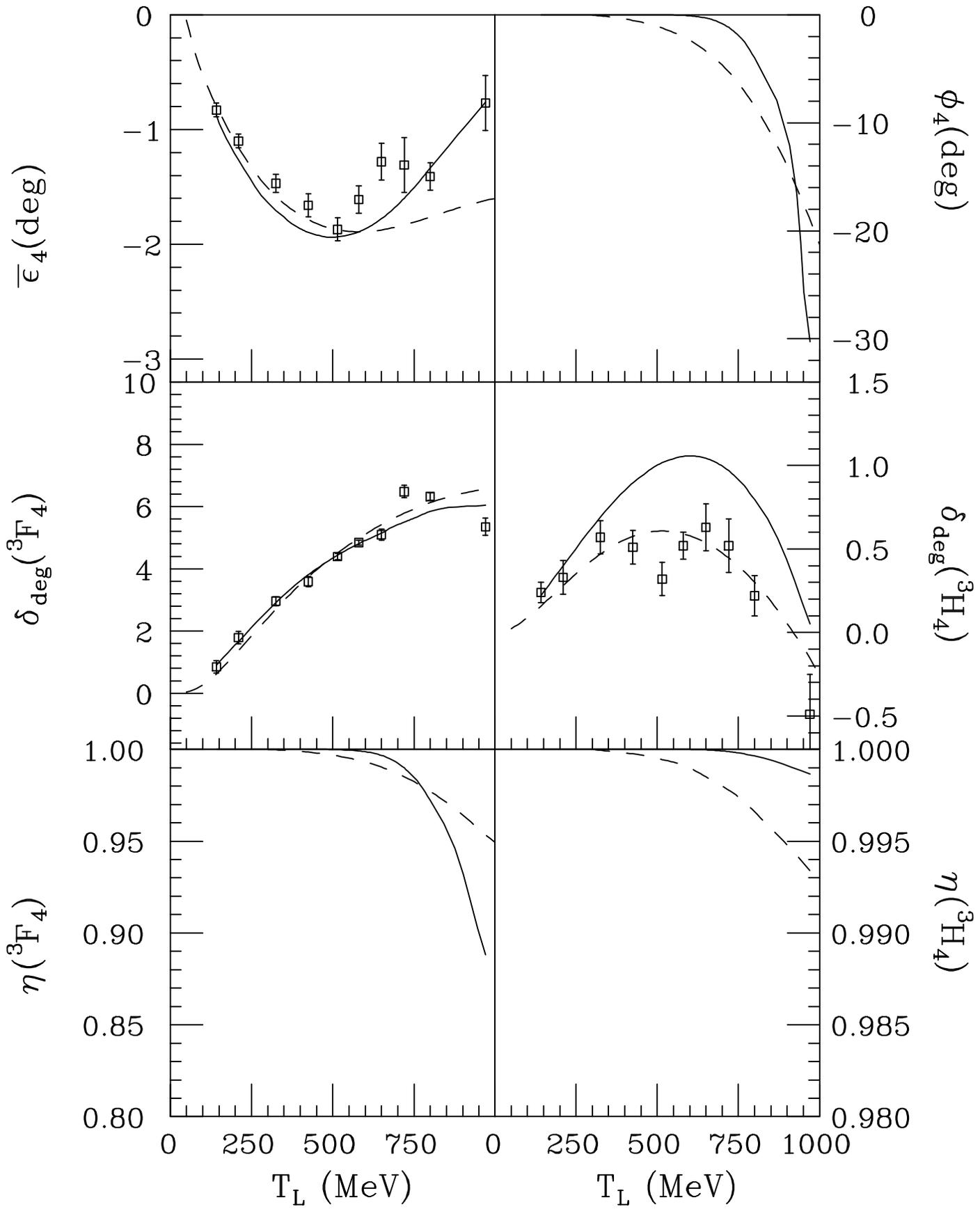
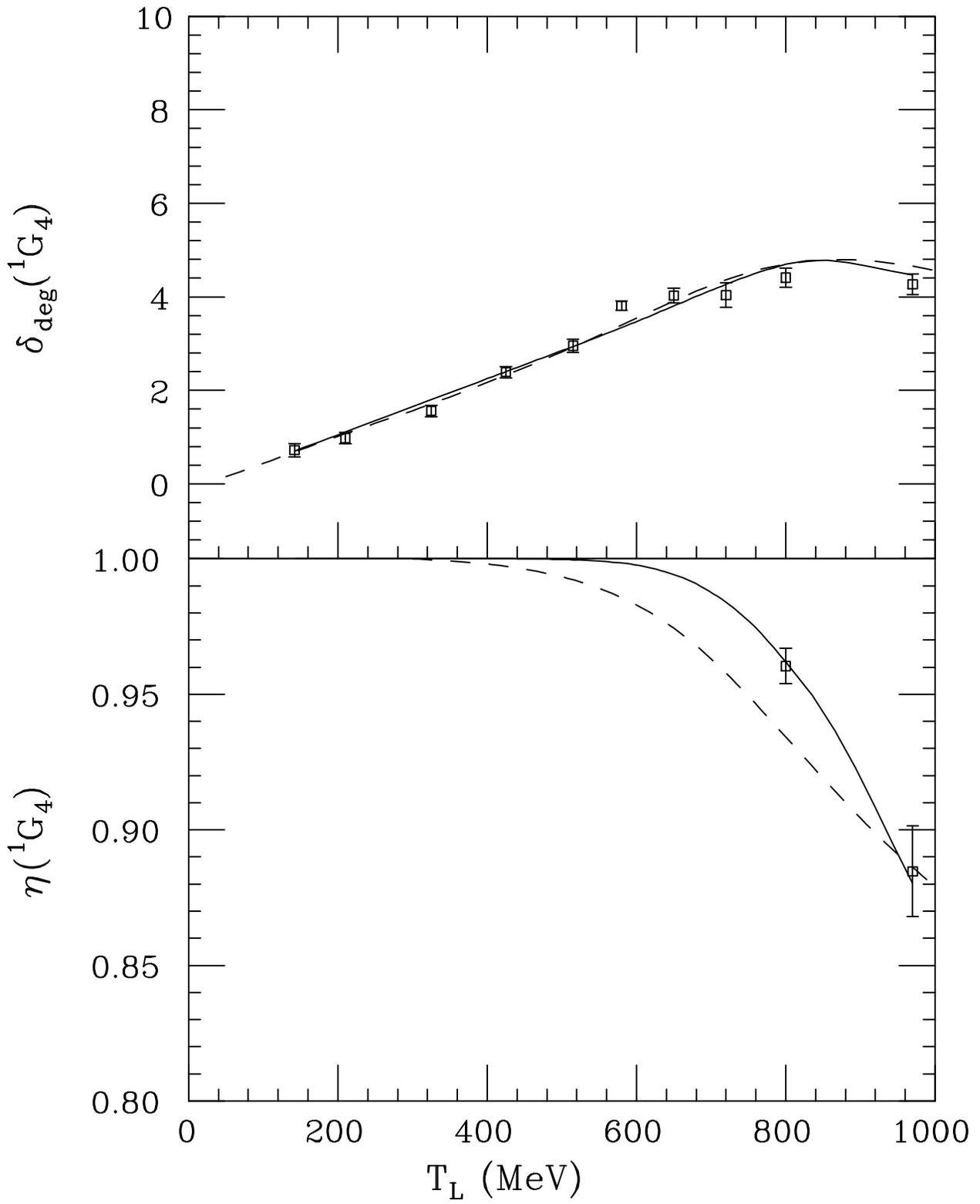
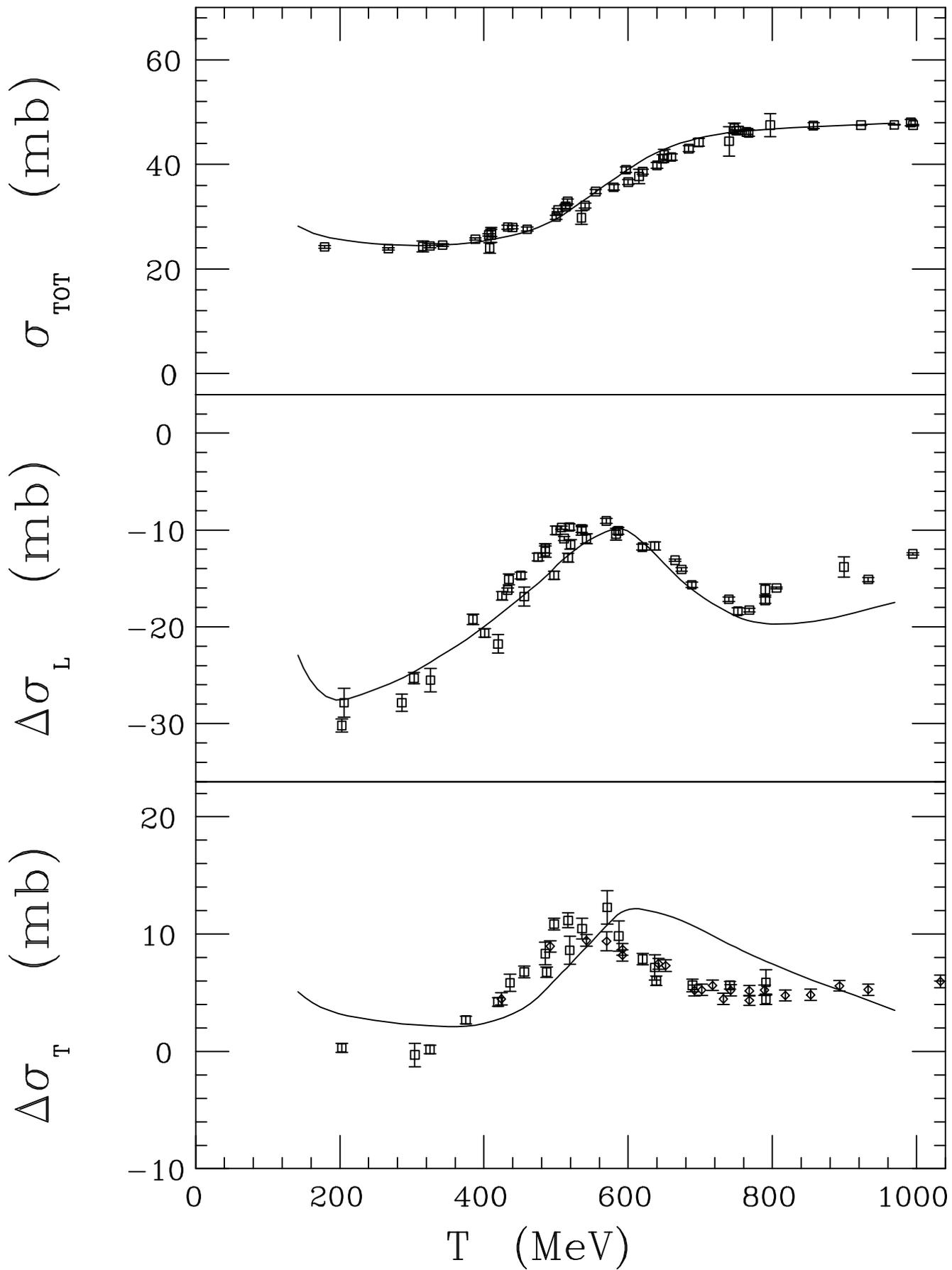
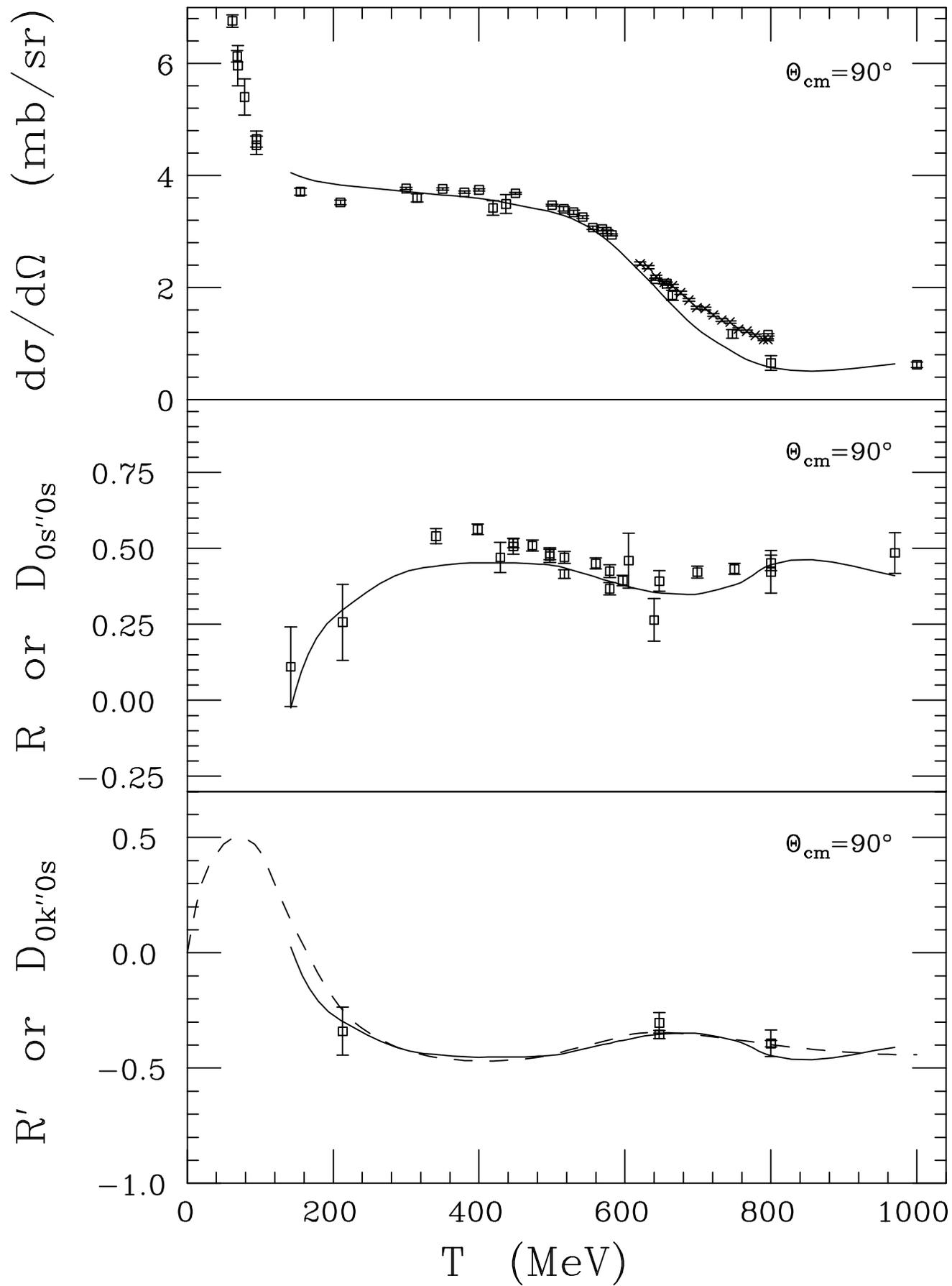
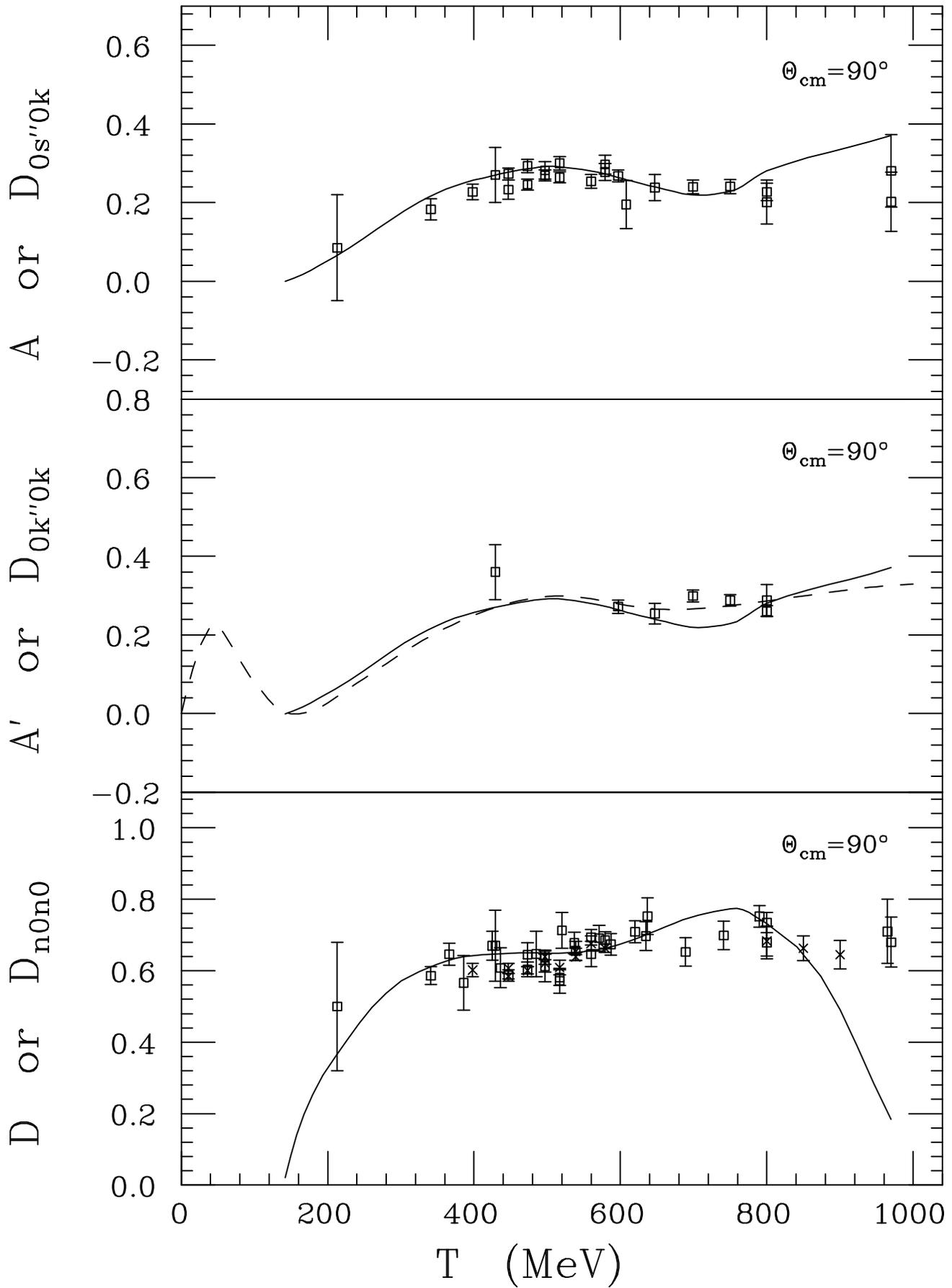
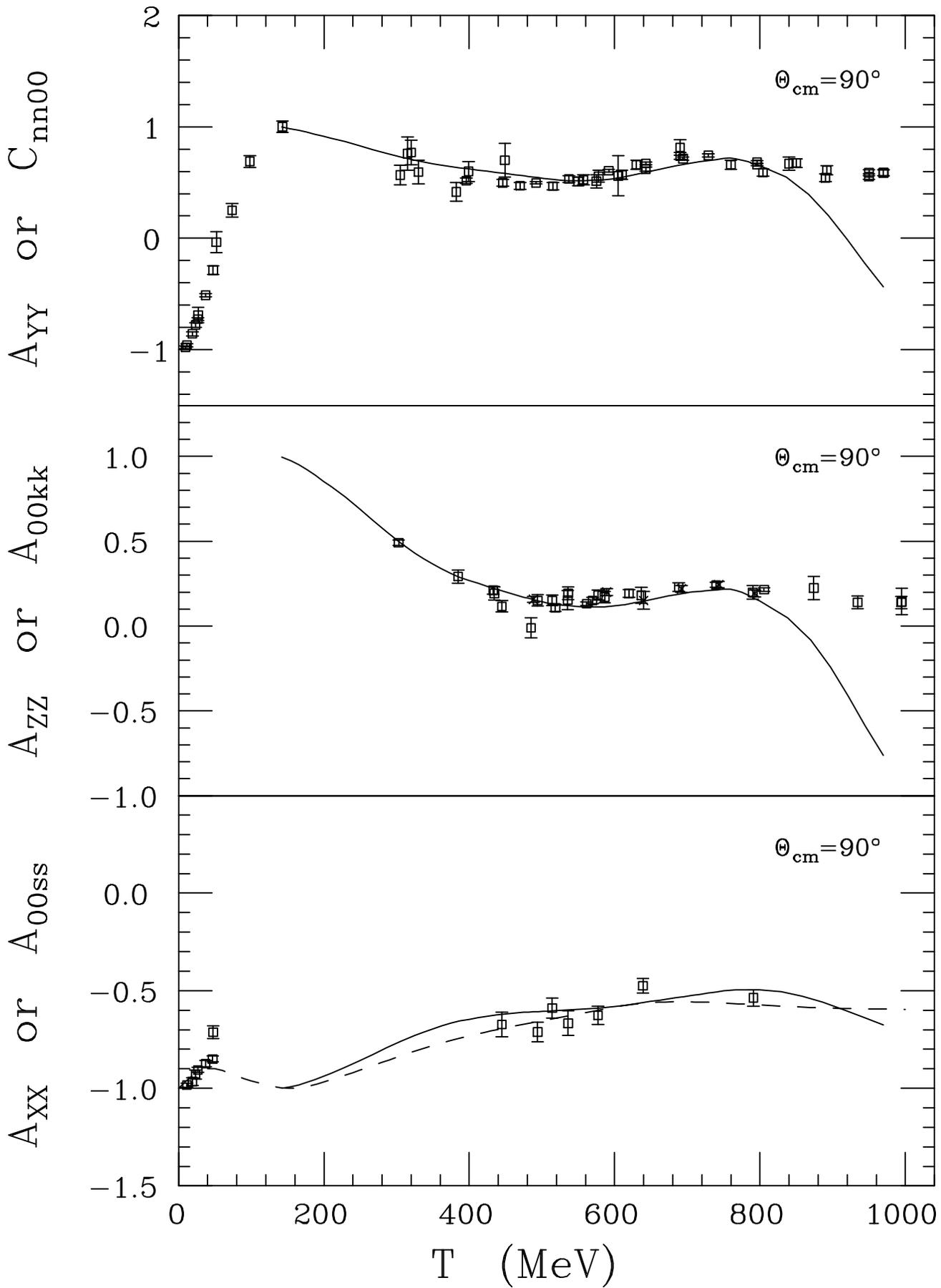
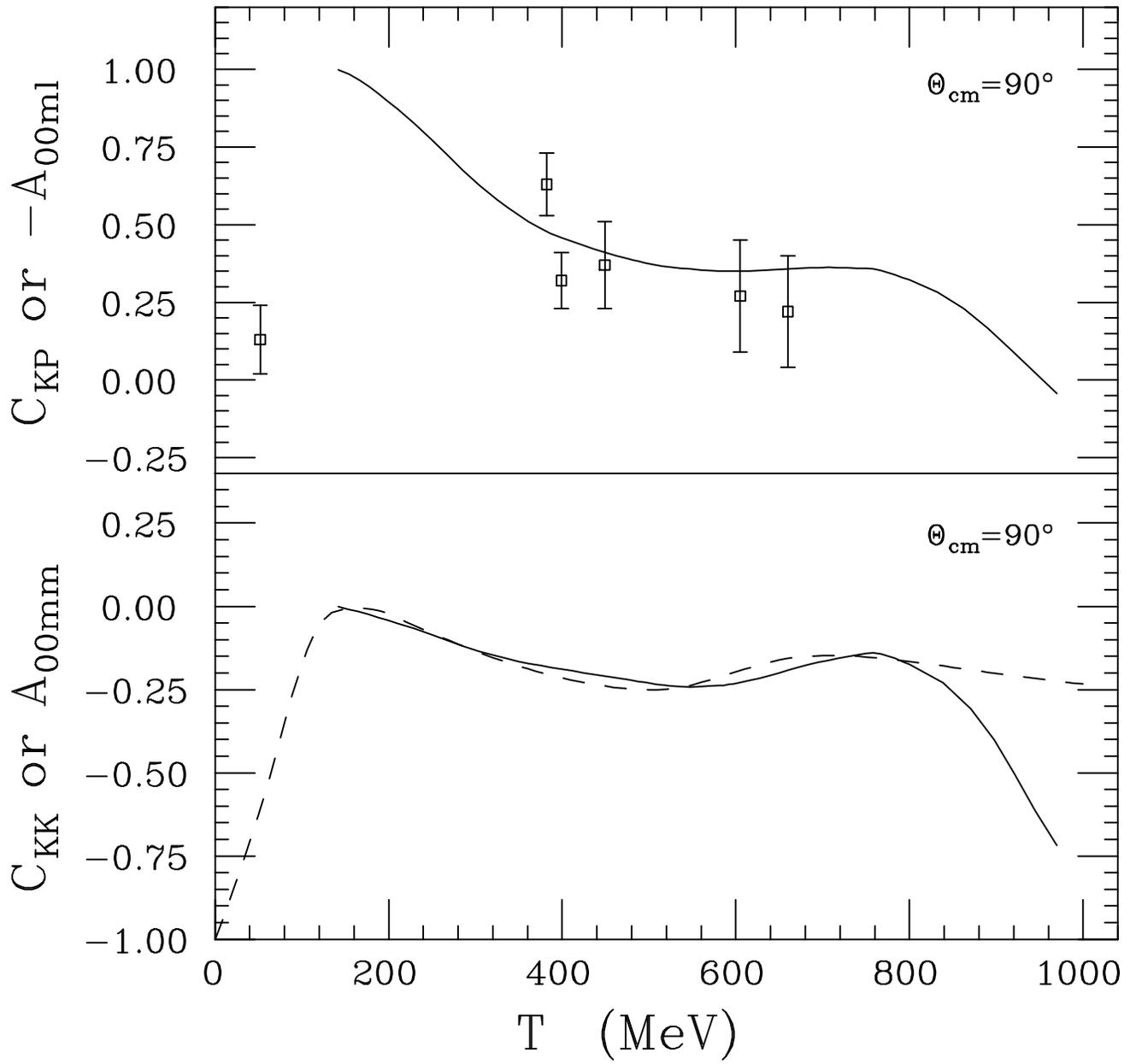
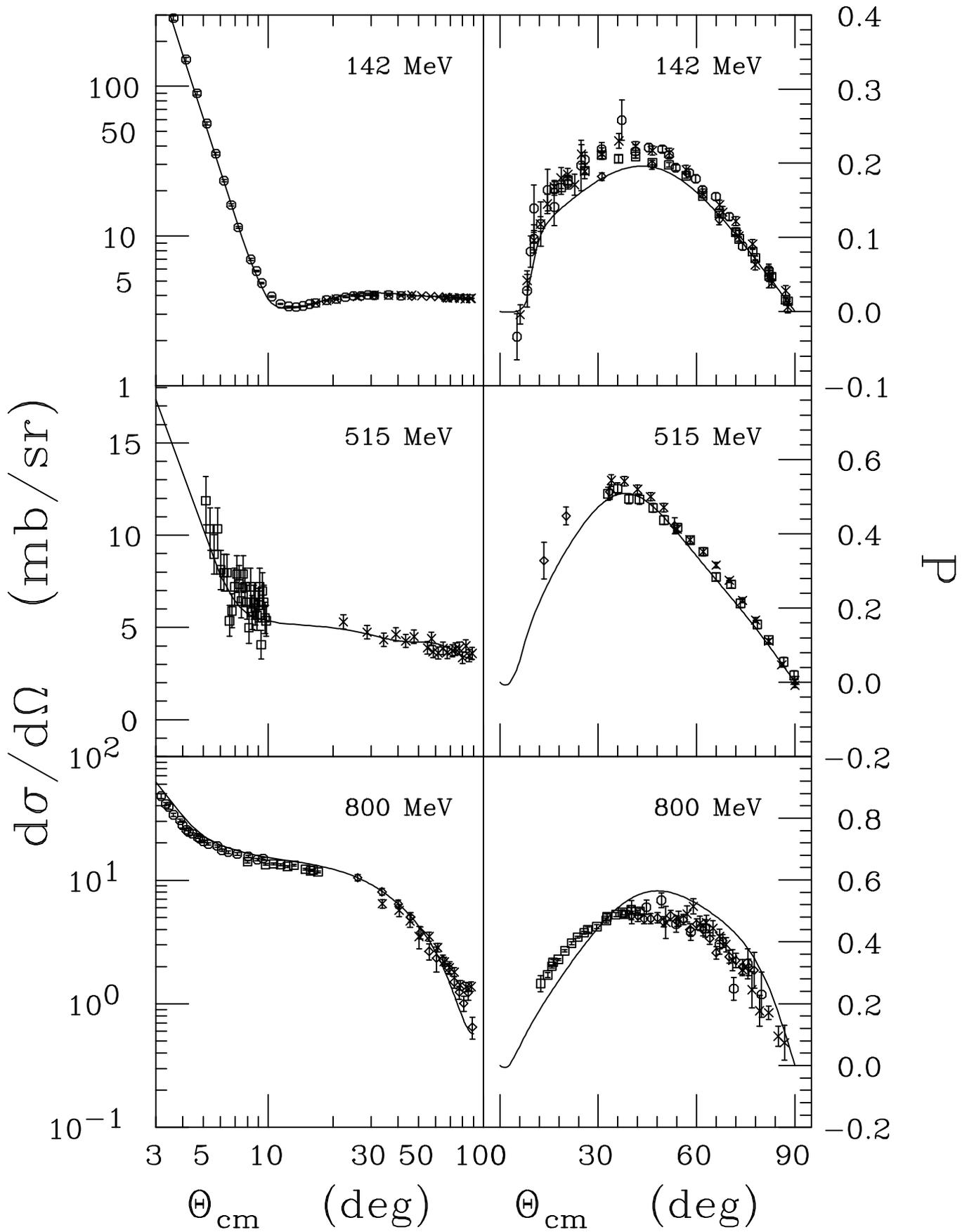
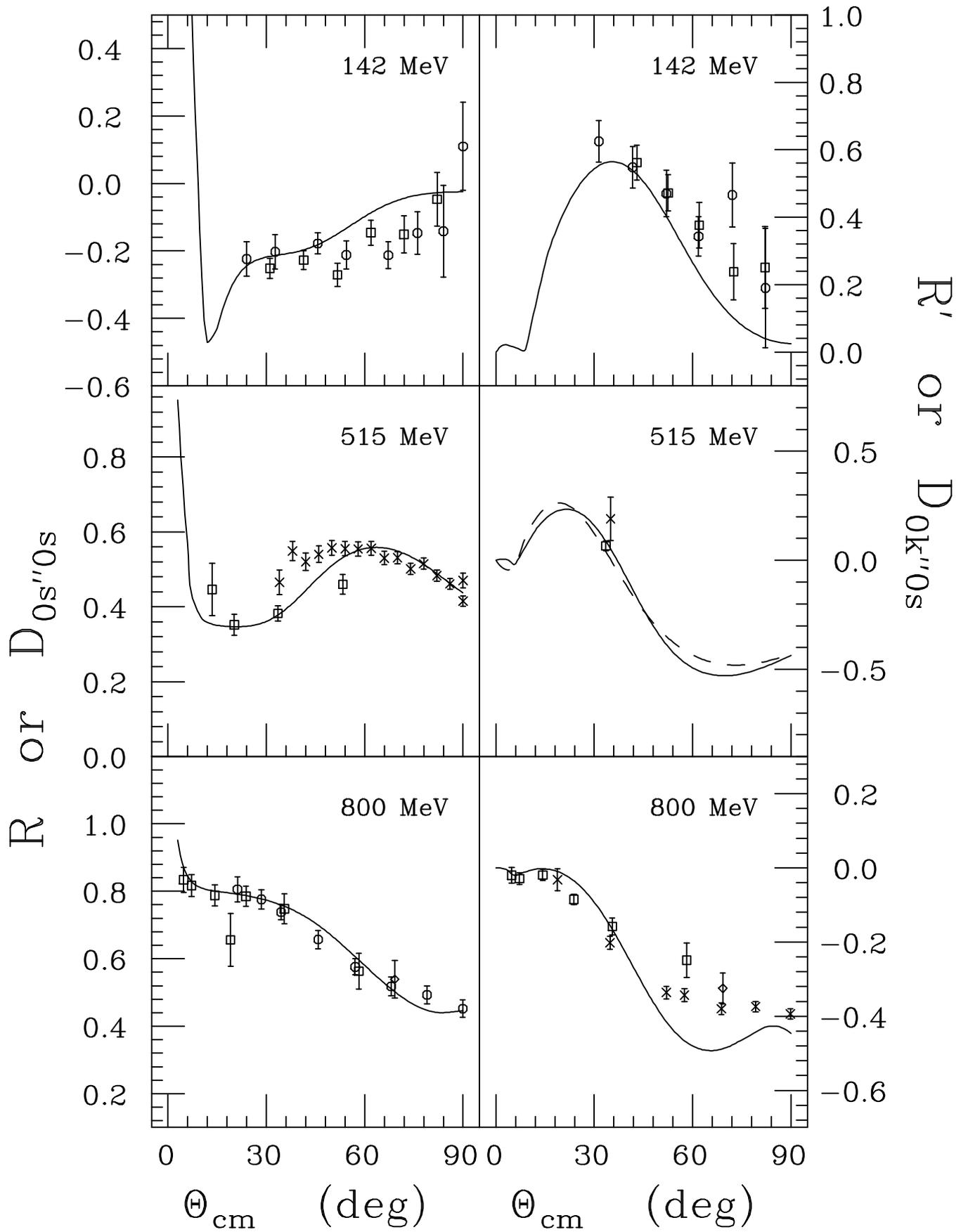
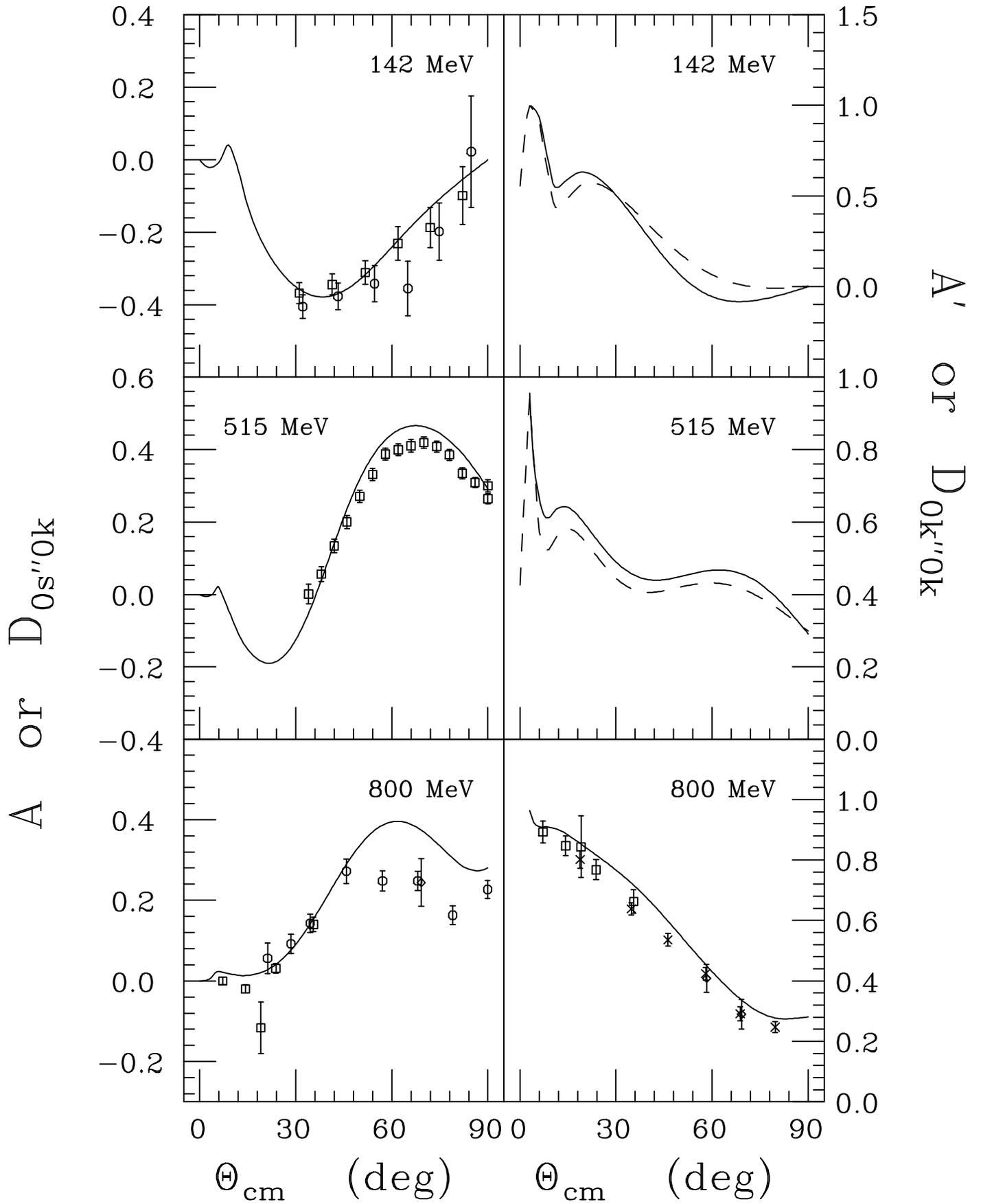
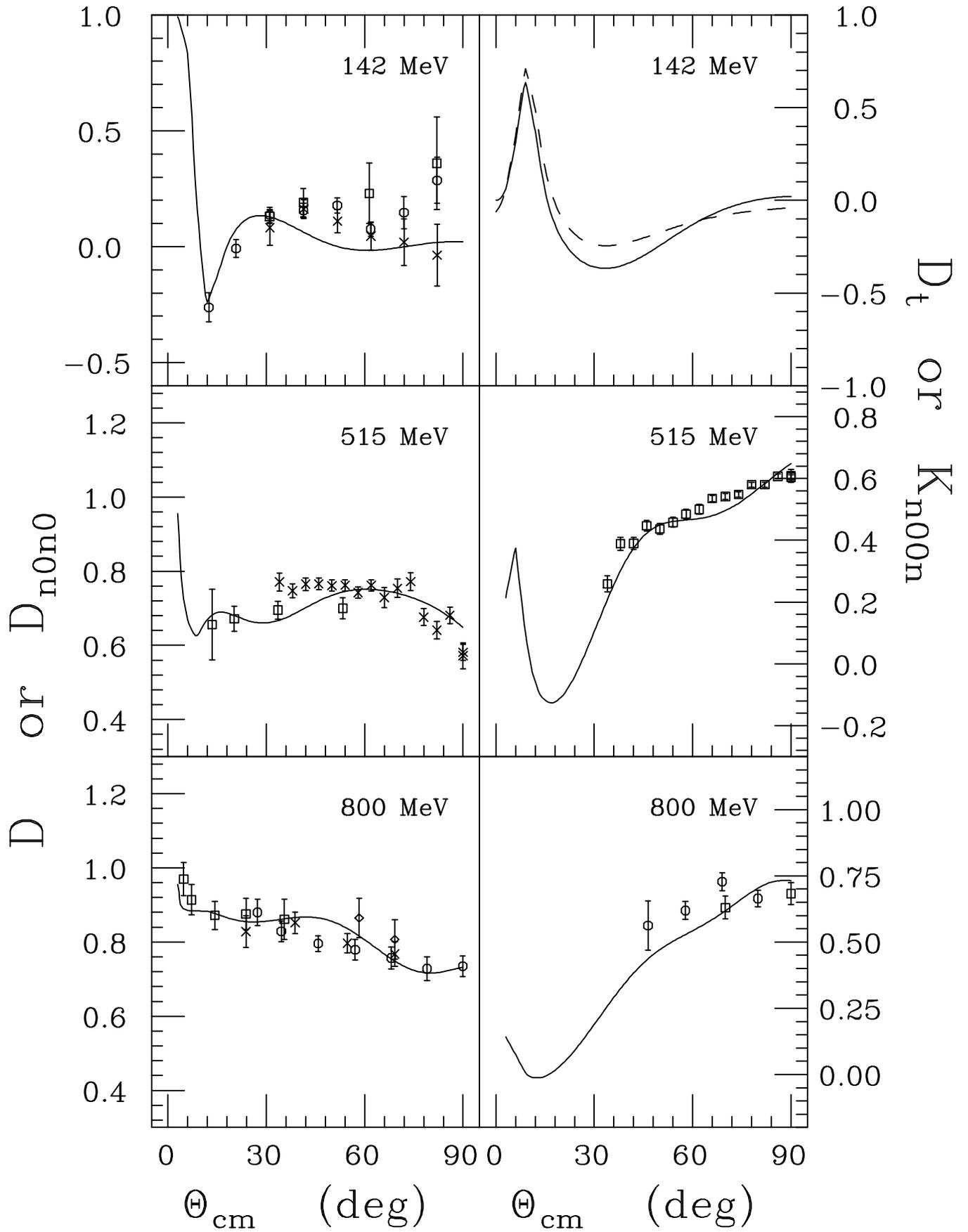
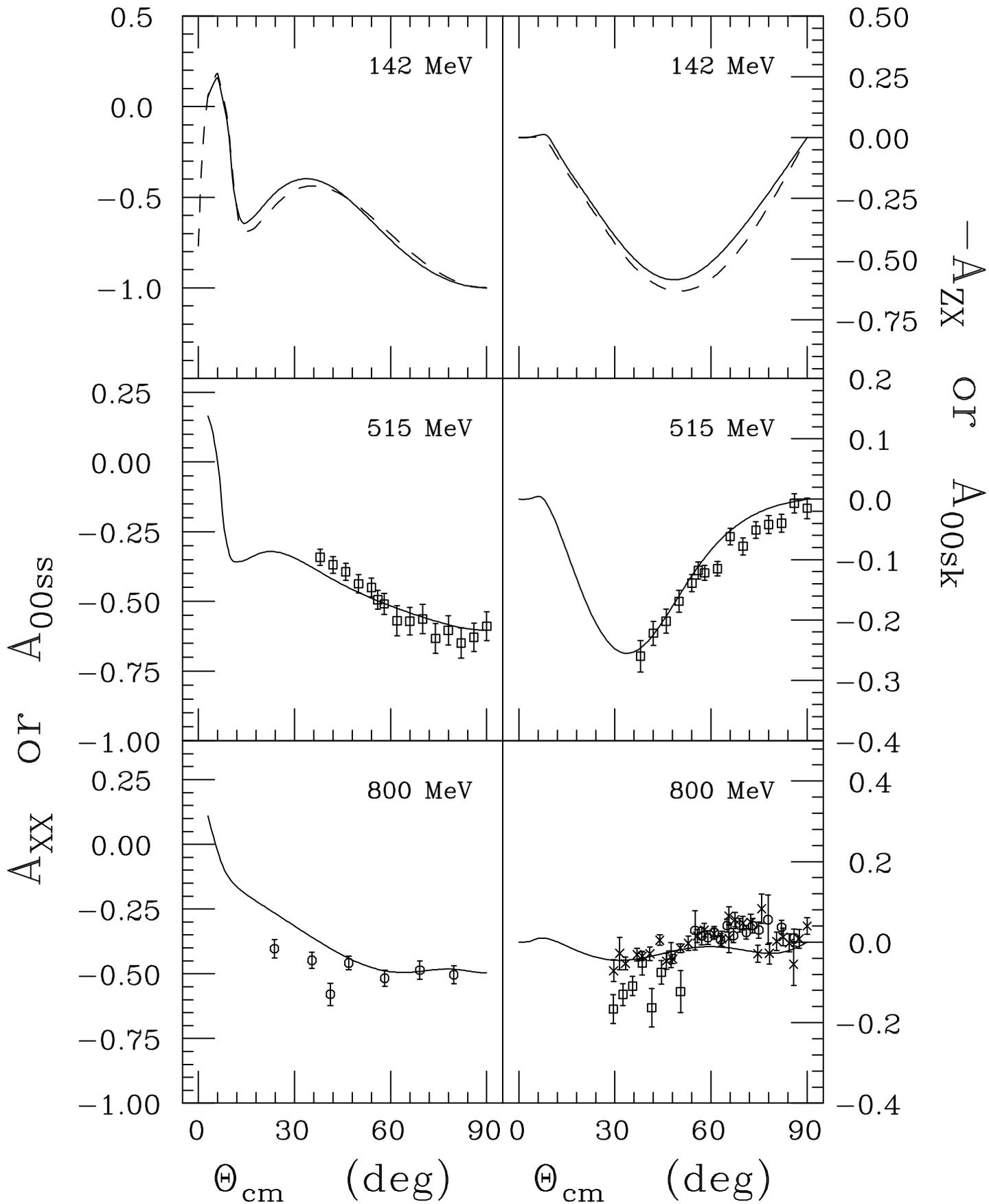
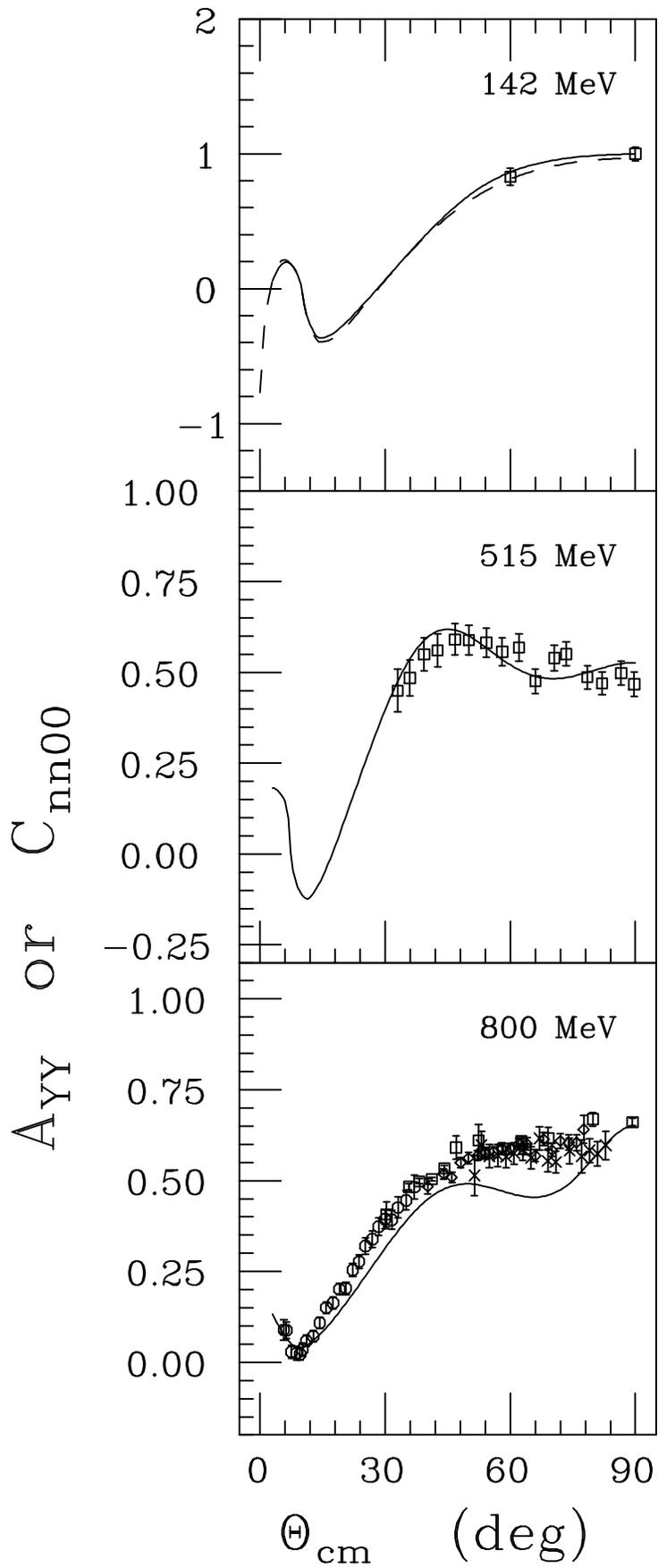
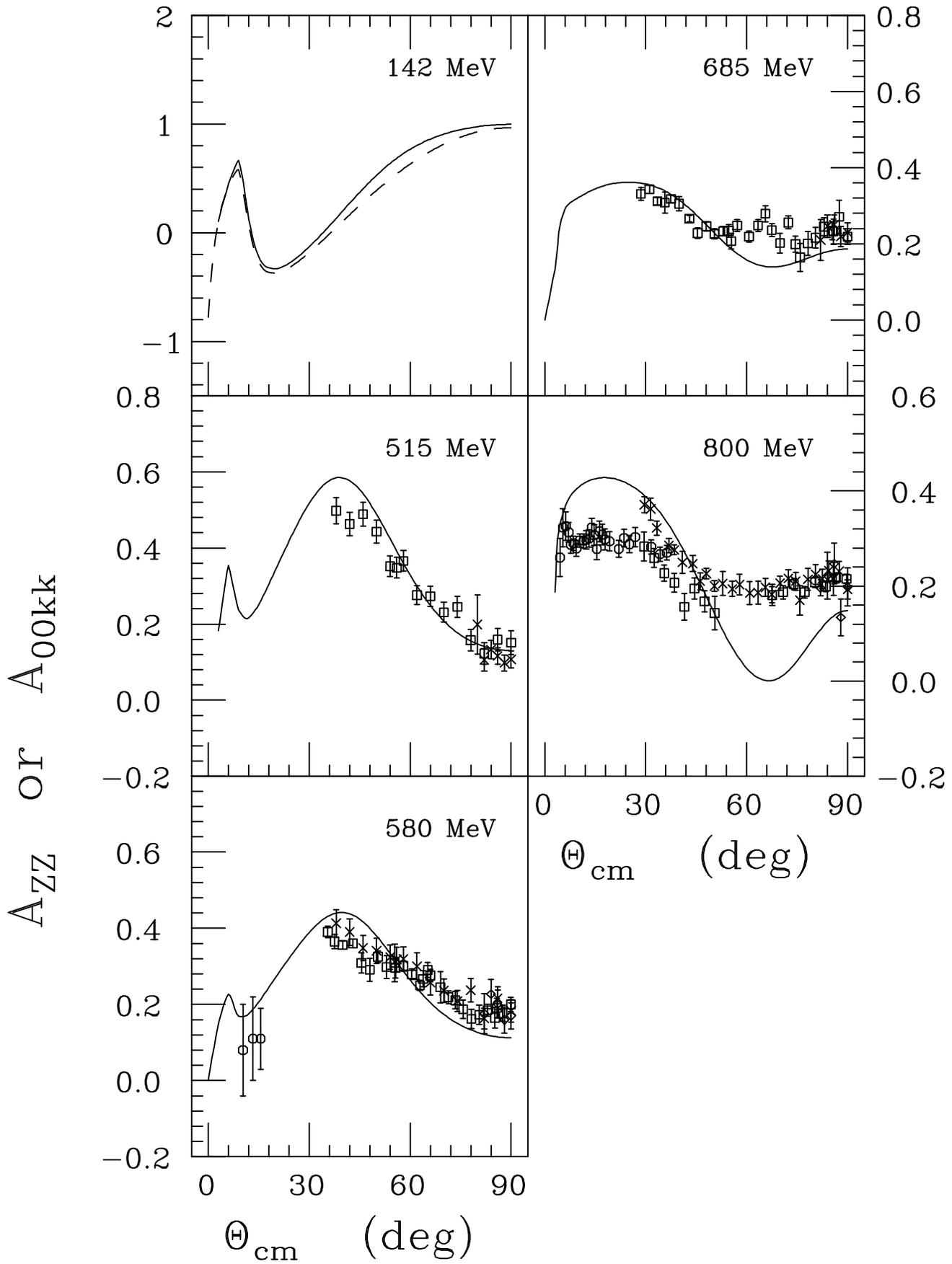
\end